\documentclass{aa} 
\usepackage{graphicx}
\usepackage[fleqn]{amsmath}
\usepackage{multirow}
\usepackage{xspace}
\usepackage{txfonts}
\usepackage{color}
\usepackage{natbib,twoopt}
\usepackage[breaklinks=true]{hyperref} 
\usepackage{float}
\usepackage[caption = false]{subfig}

\hypersetup{
  colorlinks=true,   
  citecolor=blue,    
  urlcolor=blue,     
  linkcolor=blue,    
}

\bibpunct{(}{)}{;}{a}{}{,} 

\makeatletter
\newcommandtwoopt{\citeads}[3][][]{\href{http://adsabs.harvard.edu/abs/#3}%
{\def\hyper@linkstart##1##2{}%
\let\hyper@linkend\@empty\citealp[#1][#2]{#3}}}
\newcommandtwoopt{\citepads}[3][][]{\href{http://adsabs.harvard.edu/abs/#3}%
{\def\hyper@linkstart##1##2{}%
\let\hyper@linkend\@empty\citep[#1][#2]{#3}}}
\newcommandtwoopt{\citetads}[3][][]{\href{http://adsabs.harvard.edu/abs/#3}%
{\def\hyper@linkstart##1##2{}%
\let\hyper@linkend\@empty\citet[#1][#2]{#3}}}
\newcommandtwoopt{\citeyearads}[3][][]%
{\href{http://adsabs.harvard.edu/abs/#3}
{\def\hyper@linkstart##1##2{}%
\let\hyper@linkend\@empty\citeyear[#1][#2]{#3}}}
\makeatother

    \newcommand{\kms}{\hbox{km s$^{-1}$}\xspace}
    \newcommand{\fluxcgs}{\hbox{erg s$^{-1}$ cm$^{-2}$}\xspace}

    \newcommand{\lumcgs}{\hbox{erg s$^{-1}$}\xspace}
    \newcommand{\ang}{\AA\xspace}
    \newcommand{\feka}{Fe\,K$\alpha$\xspace}

    \newcommand{\hb}{H$\beta$\xspace}
    \newcommand{\civ}{C\,\textsc{iv}\xspace}
    \newcommand{\siiv}{Si\,\textsc{iv}\xspace}
    \newcommand{\siv}{S\,\textsc{iv}\xspace}

    \newcommand{\nh}{N_{\rm H}}

    \newcommand{\lya}{Ly$\alpha$\xspace}
    
    \newcommand{\lbol}{L_{\rm bol}}
    \newcommand{\luv}{L_{\rm UV}}
    \newcommand{\fuv}{F_{\rm UV}}
    \newcommand{\lduv}{L_{\rm 2500\,\ang}}
    \newcommand{\lx}{L_{\rm X}}
    \newcommand{\fx}{F_{\rm X}}
    \newcommand{\ldx}{L_{\rm 2\,keV}}

    \newcommand{\mbh}{M_{\rm BH}}

    \newcommand{\aox}{\alpha_{\rm ox}}
    \newcommand{\daox}{\Delta\alpha_{\rm ox}}

    \newcommand{\ebv}{E(B-V)}
    \newcommand{\xmm}{\emph{XMM--Newton}\xspace}

    \newcommand{\chandra}{\emph{Chandra}\xspace}

    \newcommand{\rosat}{\emph{ROSAT}\xspace}
    \newcommand{\xspec}{\textsc{xspec}\xspace}
    \newcommand{\ciao}{\textsc{ciao}\xspace}

    \newcommand{\sas}{\textsc{sas}\xspace}

    \newcommand{\rmfgen}{\texttt{rmfgen}\xspace}
    \newcommand{\arfgen}{\texttt{arfgen}\xspace}
    
    \newcommand{\matplotlib}{\texttt{matplotlib}\xspace}
    \newcommand{\python}{\textsc{python}\xspace}
    \newcommand{\topcat}{\texttt{TOPCAT}\xspace}
    
    \newcommand{\pc}{\phantom{0}}
    \newcommand{\ps}{\phantom{00}}
    \newcommand{\pn}{\phantom{$-$}}

\begin{document} 

\title{The most luminous blue quasars at 3.0\,$<$\,\textit{z}\,$<$\,3.3}
\subtitle{I. A tale of two X-ray populations}

\author{E.~Nardini\inst{1}, 
        E.~Lusso\inst{2,1},
        G.~Risaliti\inst{2,1},
        S.~Bisogni\inst{2,1}, 
        F.~Civano\inst{3},
        M.~Elvis\inst{3},
        G.~Fabbiano\inst{3},
        R.~Gilli\inst{4},
        A.~Marconi\inst{2,1}, 
        F.~Salvestrini\inst{5,4},
        \and
        C.~Vignali\inst{5,4}
        }

\institute{
$^{1}$INAF -- Osservatorio Astrofisico di Arcetri, Largo Enrico Fermi 5, I-50125 Firenze, Italy \\
              \email{emanuele.nardini@inaf.it} \\
$^{2}$Dipartimento di Fisica e Astronomia, Universit\`a di Firenze, via G. Sansone 1, I-50019 Sesto Fiorentino, Firenze, Italy \\
$^{3}$Center for Astrophysics | Harvard \& Smithsonian, 60 Garden Street, Cambridge, MA 02138, USA \\
$^{4}$INAF -- Osservatorio di Astrofisica e Scienza dello Spazio di Bologna, via Gobetti 93/3, I-40129 Bologna, Italy\\
$^{5}$Dipartimento di Fisica e Astronomia, Universit\`a degli Studi di Bologna, via Gobetti 93/2, I-40129 Bologna, Italy
          }


 
\abstract
{We present the X-ray analysis of 30 luminous quasars at $z \simeq 3.0$--3.3 with pointed \xmm observations (28--48 ks) originally obtained by our group to test the suitability of active galactic nuclei as standard candles for cosmological studies. The sample was selected in the optical from the Sloan Digital Sky Survey Data Release 7 to be representative of the most luminous, intrinsically blue quasar population, and by construction boasts a high degree of homogeneity in terms of optical and UV properties. In the X-rays, only four sources are too faint for a detailed spectral analysis, one of which is formally undetected. Neglecting one more object later found to be radio-loud, the other 25 quasars are, as a whole, the most X-ray luminous ever observed, with rest-frame 2--10 keV luminosities of 0.5--$7 \times 10^{45}$ \lumcgs. The continuum photon index distribution, centred at $\Gamma \sim 1.85$, is in excellent agreement with those in place at lower redshift, luminosity, and black-hole mass, confirming the universal nature of the X-ray emission mechanism in quasars. Even so, when compared against the well-known $\lx$--$\luv$ correlation, our quasars show an unexpectedly varied behaviour, splitting into two distinct subsets. About two-thirds of the sources are clustered around the relation with a minimal scatter of 0.1 dex, while the remaining one-third appear to be X-ray underluminous by factors of $>$\,3--10. Such a large incidence ($\approx$25\%) of X-ray weakness has never been reported in radio-quiet, non-broad absorption line (BAL) quasar samples. Several factors could contribute to enhancing the X-ray weakness fraction among our $z\simeq3$ blue quasars, including variability, mild X-ray obscuration, contamination from weak-line quasars, and missed BALs. However, the X-ray weak objects also have, on average, flatter spectra, with no clear evidence of absorption. Indeed, column densities in excess of a few $\times$10$^{22}$ cm$^{-2}$ can be ruled out for most of the sample. We suggest that, at least in some of our X-ray weak quasars, the corona might experience a radiatively inefficient phase due to the presence of a powerful accretion-disc wind, which substantially reduces the accretion rate through the inner disc and therefore also the availability of seed photons for Compton up-scattering. The origin of the deviations from the $\lx$--$\luv$ relation will be further investigated in a series of future studies.
} 


\titlerunning{The most luminous blue quasars at $z \simeq 3$: a tale of two X-ray populations}
\authorrunning{E. Nardini et al.}
\maketitle
%
\section{Introduction}
As the most luminous among the persistent energy sources in our Universe, quasars inevitably hold an extraordinary potential as cosmological probes. Indeed, over the last four decades, several techniques employing empirical correlations between various quasar properties have been proposed to assess the cosmological parameters. Some remarkable examples include the relations between luminosity and, in turn, emission-line equivalent width \citep{Baldwin+78}, broad-line region radius \citep{Watson+11}, and X-ray variability amplitude \citep{LaFranca+14}. However, all these correlations are either affected by large observational scatter (up to 0.6 dex) or applicable over a limited redshift range with the current facilities. Other methods (e.g. \citealt{Elvis-Karovska_02,Wang+13,Marziani-Sulentic_14}) could be promising, but are still more a proof of concept than a real cosmological tool. For these reasons, quasars (or, in general, active galactic nuclei; AGNs) are not yet competitive against standard probes like type Ia supernovae (SNe).

Further consideration must be given to the so-called $\lx$--$\luv$ relation. The X-ray and optical or ultraviolet (UV) luminosities of quasars have long been known to follow a non-linear relation (e.g.~\citealt{Avni-Tananbaum_86}), whereby optically luminous objects are relatively underluminous in the X-rays: an increase by an order of magnitude in UV luminosity typically corresponds to an increase by only a factor of four in X-ray luminosity. Through the use of fluxes, the non-linear nature of this relation can, in principle, provide a direct measure of the luminosity distance, thus turning quasars into a new class of standardizable candle. Until recently however, the dispersion of 0.35--0.40 dex (e.g.~\citealt{Vignali+03,Strateva+05,Steffen+06,Just+07,Lusso+10}) had deterred and/or undermined any attempts to use the relation for precision cosmology. After \citet{RL15}, who built the first quasar Hubble diagram (see also \citealt{Bahcall-Hills_73, Setti-Woltjer_73}) based on UV and X-ray fluxes, it has become clear that most of the observed dispersion is not intrinsic to the relation itself but is due to observational issues, among which X-ray absorption, UV extinction by dust, background filtering and calibration uncertainties in the X-rays, variability, and selection biases associated with the flux limits of the different samples. Indeed, with an optimal selection of `clean' sources, the dispersion drops to about 0.2 dex \citep{LR16,LR17}. In addition to the cosmological merit, this has the major physical consequence that a universal mechanism must be regulating the production of the optical, UV, and X-ray emission in quasars.

A deeper understanding of the X-ray--UV correlation and of its grounds is therefore mandatory to consolidate and fully exploit the potential of quasars in cosmology. This is chiefly a problem of supermassive black hole (SMBH) accretion physics, which can only be addressed by investigating the intimate connection between the accretion disc and the enigmatic X-ray corona. Yet, even under the most simplistic assumptions (e.g. a geometrically thin, optically thick disc; \citealt{SS73}), testing any model of accretion-driven emission with AGN spectra is far from trivial. The main limitation, which exceeds the capabilities of a single observatory, is the necessity of exhaustive, ideally simultaneous spectral information to properly determine the shape of the relevant part (i.e.~optical to X-rays) of the spectral energy distribution (SED). Moreover, the peak of the emission from the accretion disc (the `big blue bump'; \citealt{Czerny-Elvis_87}) should be adequately probed. As most AGN spectra roll over in the extreme UV, this range is only accessible to ground-based facilities in high-redshift objects, at the cost of a complete lack \citep{Capellupo+16} or modest quality \citep{Collinson+17} of the corresponding X-ray spectra. 

Many of these difficulties can be overcome through a tailored selection. A promising sample in this respect has been used in the recent work by \citet{RL19}, to which we refer for the intriguing implications on the cosmological side. The sample consists of 30 quasars at $z\simeq3$ with pointed \xmm observations, specifically designed to fill the quasar Hubble diagram in a redshift range that is still unexplored by means of SNe \citep{Scolnic+18} and baryon acoustic oscillations \citep{Blomqvist+19}. This is the first in a series of papers dedicated to the study of the physical properties of these 30 quasars, for which we have been assembling a valuable broadband coverage over the past few years. The paper is organised as follows: in Section~\ref{sel} we describe the selection of the sample, while Section~\ref{obs} is dedicated to the X-ray observations and data reduction. The spectral analysis is presented in Section~\ref{spe}, and the results are discussed in Section~\ref{dis}. Conclusions are drawn in Section~\ref{con}.

\begin{figure}
\centering
\includegraphics[width=8.5cm]{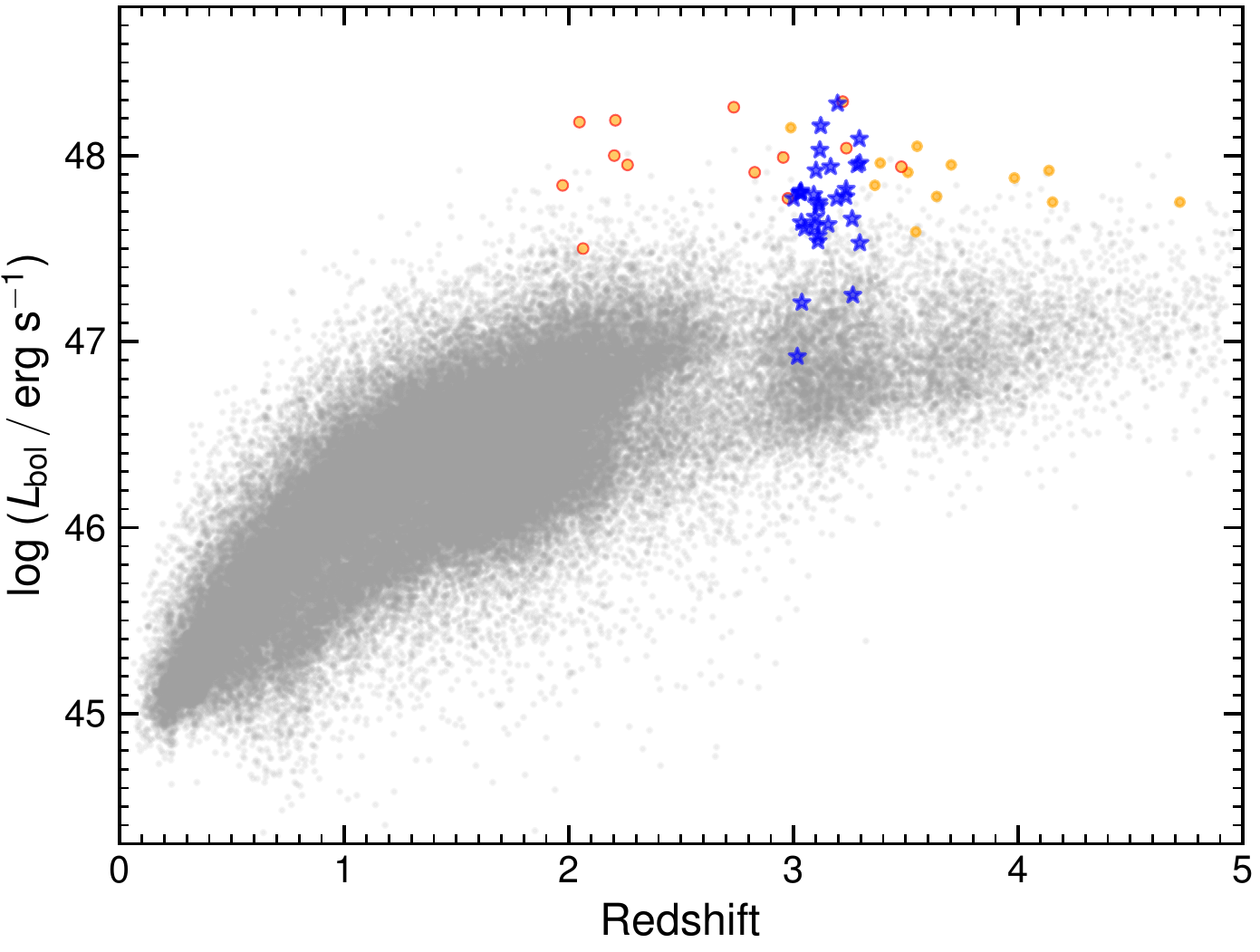}
\caption{Location of the 30 quasars at $z \simeq 3.0$--3.3 analysed in this work (blue stars) in the luminosity--redshift plane of the SDSS-DR7 catalogue (\citealt{Shen+11}), where radio-loud and BAL quasars have been removed (91,484 objects, grey dots). Also shown for comparison, with the same filters applied, are the remaining 25 sources from the WISSH sample (6 more are in common) with available X-ray data (in orange, from \citealt{Martocchia+17}; the objects for which the number of counts was enough to perform a spectral analysis are flagged with red circles).}
\label{zl}
\end{figure}

\section{Sample selection}
\label{sel}
The present work is based on an extensive X-ray campaign performed with \xmm (cycle 16, proposal ID: 080395, PI: G.\,Risaliti), which observed 30 quasars in the $z\simeq 3.0$--3.3 redshift range for a total exposure of 1.13 Ms. The targets were selected from the catalogue of quasar properties of \citet{Shen+11}, drawn from the Sloan Digital Sky Survey (SDSS) seventh Data Release (DR7) and consisting of 105,783 spectroscopically confirmed broad-line quasars. From this catalogue, we removed all the entries flagged as either `broad absorption line' (BAL; 6214 sources with BAL flag $>$\,0) or `radio-loud' (8257 sources with radio-loudness parameter $R=F_{\nu,\,6\,\rm{cm}}/F_{\nu,\,2500\,\ang}\geq10$).\footnote{There are 420 objects that meet both criteria.} Such a selection yields a pre-cleaned sample of 91,732 SDSS quasars, from which we further excluded 136 objects classified as BALs by \citet{Gibson+09} and 94 objects with no measure of the bolometric luminosity in the catalogue.\footnote{Depending on redshift, bolometric luminosities are computed from one of the 5100, 3000, or 1350\,\ang monochromatic luminosities \citep{Shen+11}, using the bolometric corrections of \citet{Richards+06}.} Subsequently, we checked for additional radio-loud sources against the MIXR sample, which is~the largest available mid-infrared (WISE), X-ray (3XMM), and radio (FIRST+NVSS) collection of AGNs and star-forming galaxies \citep{Mingo+16}. Of the remaining SDSS-DR7 quasars, 18 fall within a matching radius of 2$\arcsec$ from one of the 918/2753 MIXR objects that are considered to be radio-loud based on multiwavelength diagnostics, and were therefore neglected. This leads to a clean parent sample of 91,484 quasars, whose distribution of bolometric luminosity as a function of redshift is shown in Figure~\ref{zl} (grey dots).

Moving from here, we applied several other filters to define a homogenous quasar sample around $z \sim 3$, the highest redshift for which an X-ray spectrum of good quality can be obtained with a reasonable exposure (a few tens of ks). Specifically: 1) we first restricted the sample to the narrow redshift range $3.0 < z < 3.3$, which is populated by 2566 quasars. 2) We then selected all the sources (1005) with an estimated bolometric luminosity in excess of $8\times10^{46}$ \lumcgs. 3) Following the approach described in \citet{LR16}, we singled out a sub-sample of objects where intrinsic reddening is small. In short, we built for each quasar a broadband SED using the available photometry from several surveys, from the rest-frame UV (SDSS) to the near-infrared (i.e.~2MASS, WISE). We computed the slopes of a $\log(\nu)$--$\log(\nu L_{\nu})$ power law in the 0.3--1\,$\mu$m ($\Gamma_1$) and 1450--3000\,\ang ($\Gamma_2$) rest-frame bands, and retained only the sources (about 70\%) with $\Gamma_1-\Gamma_2$ centred at $\ebv = 0.0$ with a radius of 1.1, which roughly corresponds to $\ebv \simeq 0.1$. 4) Finally, we sorted the surviving objects by brightness at rest-frame 2500\,\ang, where the observed UV flux density ($F_{2500\,\ang}$, in \fluxcgs Hz$^{-1}$) is provided in the \citet{Shen+11} catalogue through a power-law continuum fit to the SDSS spectrum.\footnote{Host contamination at 2500\,\ang is negligible at high luminosity.} We chose the top 30 quasars with optimal \xmm observing conditions (e.g.~visibility, low Galactic column). The final sample, introduced in Table~\ref{st} and represented by blue stars in Figure~\ref{zl}, includes a fraction of objects for which there are existing X-ray snapshots (a few ks; see Section~\ref{dis}), which supported the feasibility of our X-ray follow-up campaign.

The filters above were primarily designed to identify a subset of quasars with uniform UV properties, such as continuum luminosity and spectral slope. Since none of the selected objects are included in the MIXR catalogue, before the analysis we independently recomputed the radio-loudness parameter to verify that all 30 quasars in the $z\simeq3$ sample have negligible radio emission. This is confirmed for every source but one, SDSS~J090033.50+421547.0, which was classified as radio-quiet ($R\simeq2$) in the SDSS-DR7 but in fact turns out to be radio-loud. An integrated flux density of 1.71 mJy (peak flux of 1.52 mJy/beam) at 1.4 GHz was obtained for J0900+42 from a cross-match with the FIRST survey, which, under the same assumption of \citet{Shen+11} of a power law with a slope of $-0.5$ to estimate the rest-frame flux density at 6 cm, leads to $R\gg10$. We therefore flag J0900+42 at each step of the analysis, excluding it from any general consideration regarding the sample. 

In keeping with the selection criteria, the quasars in the $z\simeq3$ sample have indeed highly homogenous UV spectra, characterised by an intrinsically blue continuum. This is demonstrated by their average spectral stack, which was built following the procedure described in \citet{Lusso+15}. Briefly, we took into account all the SDSS spectra of the 29 radio-quiet sources, including multiple observations when available (37 SDSS spectra), and corrected the observed flux density for Galactic reddening by adopting the $\ebv$ values from \citet{Schlegel+98}\footnote{The median reddening is $\ebv=0.03$ mag.} and the Galactic extinction curve from \citet{Fitzpatrick99}, with $R_V=3.0$. We then shifted each quasar spectrum to the rest frame and linearly interpolated over a rest-frame wavelength array with fixed dispersion, $\Delta\lambda\simeq0.3$\,\ang,\footnote{This is half of the ratio $1250\,\ang/R_\lambda$, where we have assumed a spectral resolution of $R_\lambda = 2000$ at 1250 \ang.} normalizing to the 1450\,\ang~flux. All the flux values were finally averaged to produce the stacked spectrum, rescaled to unity at $\lambda=1450$\,\ang. Uncertainties were estimated through a bootstrap resampling technique, creating 5000 random samplings of the 37 spectra with replacement, and applying the same procedure outlined above. 

\begin{figure}
\centering
\includegraphics[width=8.5cm]{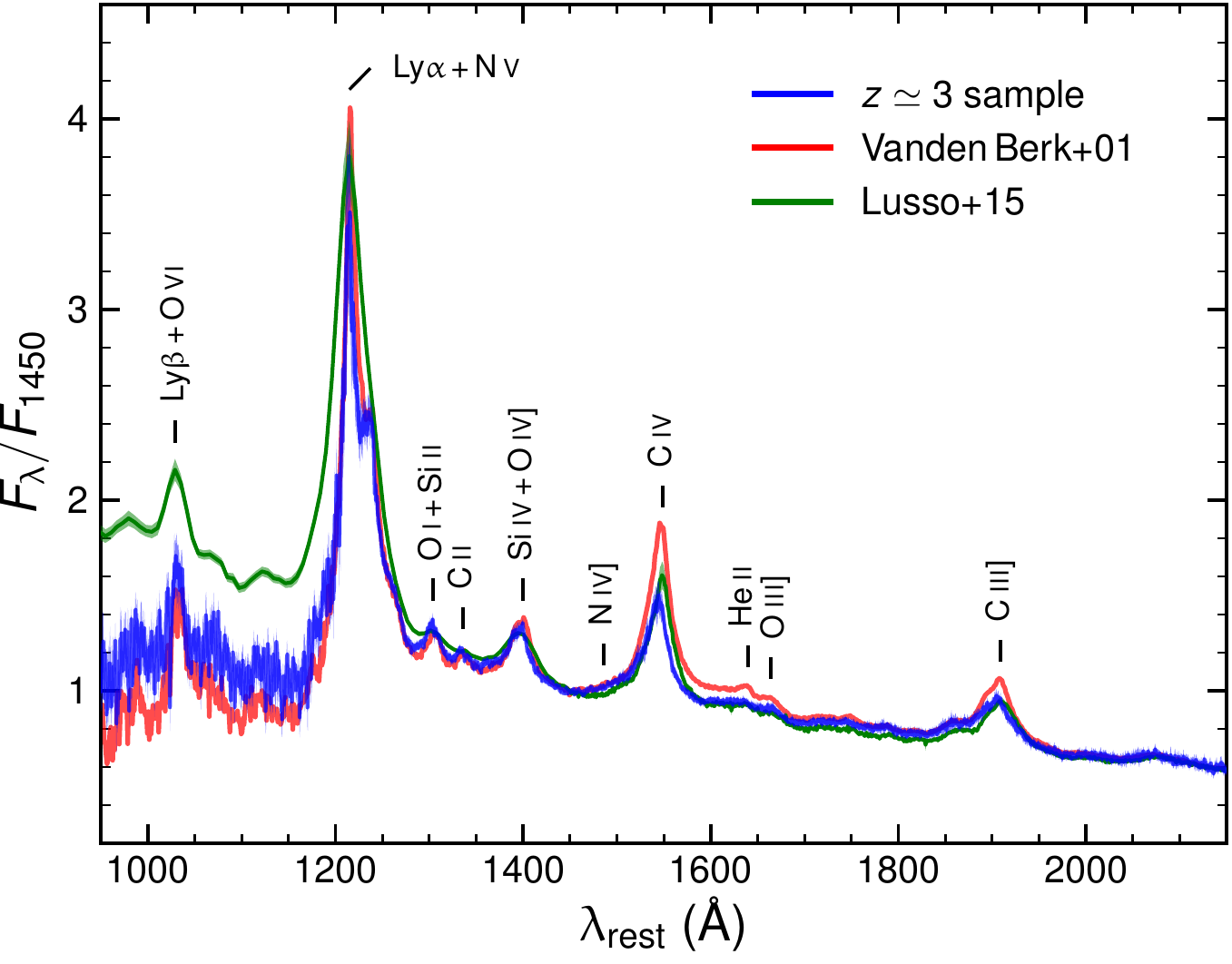}
\caption{Average UV spectrum of our $z\simeq3$ quasars as obtained from the stack of 37 SDSS observations (blue solid line; J0900+42 has been excluded). The spectrum was corrected for Galactic reddening and normalized to unity at 1450\,\ang (see Section~\ref{sel}). For comparison, we also plot the SDSS AGN composite of \citet[red solid line]{VandenBerk+01} and the stacked spectrum of $z\sim2.4$ quasars of \citet[green solid line]{Lusso+15}, corrected for IGM absorption shortwards of \lya.}
\label{as}
\end{figure}

The resulting stack is shown as the blue solid line in Figure~\ref{as}, where the small associated uncertainties are plotted as a shaded area. The average SDSS spectrum of our $z\simeq3$ quasars is compared with the AGN composite of \citet{VandenBerk+01} from the SDSS, and the one of \citet{Lusso+15} based on 53 quasars at $z\sim2.4$ and corrected for intervening absorption by neutral hydrogen in the intergalactic medium (IGM). Apart from the slight decrease in the emission-line strength, which is likely due to the Baldwin effect \citep{Baldwin77}, it is clear that both the ionizing continuum and the overall spectral properties of our sources are in very good agreement with the expected intrinsic quasar spectrum, as established by several independent works. A comprehensive UV analysis of the $z\simeq3$ sample is the subject of a forthcoming companion paper (Lusso et al., in preparation).

It is worth pointing out that our \xmm sample, when compared to other high-luminosity, high-redshift quasar compilations, is unique for several reasons. By selection, we have assembled a clear-cut and uniform subset with statistically meaningful size and excellent wavelength coverage, which can deliver a snapshot of the intrinsic quasar properties at a specific cosmic epoch (spanning only 0.2 Gyr). For example, at the same luminosity ($\lbol \sim 10^{47}$--$10^{48}$ \lumcgs; Figure~\ref{zl}) the WISE/SDSS selected hyper-luminous (WISSH; \citealt{Bischetti+17}) quasar sample boasts larger numbers (86 sources), but spreads over a much wider redshift range ($z\sim1.8$--4.6) and contains a sizeable fraction of radio-loud, BAL, and extremely red objects. Besides the cosmological value, ours is therefore an optimal sample to shed new light on the physics governing SMBH accretion in quasars and the origin of the $\lx$/$\luv$ correlation. 

\begin{table*}
\caption{Basic properties of the 30 sources in our sample and \xmm observation log.}
\label{st}
\centering
\begin{tabular}{lcccccccc}
\hline \hline \noalign{\smallskip}
SDSS name (J-) & $z$ & log\,$\lbol$ & log\,$\lduv$ & Obs. start (UTC) & Exp.\,(ks) & 
\multicolumn{3}{c}{Counts (pn, MOS\,1, MOS\,2)} \\
\noalign{\smallskip}
\hline \noalign{\smallskip}
030341.04$-$002321.9 & 3.235 & 47.82 & 31.96 & 2017/08/25\,--\,15:30:07 & 
20.4\,+\,65.9 & \pc373$\pm$24 & 161$\pm$16 & 169$\pm$16 \\
030449.85$-$000813.4 & 3.296 & 47.96 & 32.01 & 2017/08/26\,--\,04:03:26 &
30.6\,+\,72.4 & \pc659$\pm$28 & 212$\pm$17 & 221$\pm$17 \\
082619.70$+$314847.9 & 3.098 & 47.67 & 31.89 & 2017/10/26\,--\,11:13:45 & 
10.2\,+\,35.4 & \ps72$\pm$13 & \pc28$\pm$\pc9 & \pc55$\pm$\pc9 \\
083535.69$+$212240.1 & 3.110 & 47.54 & 31.91 & 2017/04/12\,--\,04:03:35 & 
22.9\,+\,74.5 & \pc492$\pm$27 & 182$\pm$16 & 235$\pm$17 \\
090033.50$+$421547.0 & 3.294 & 48.09 & 32.33 & 2017/11/17\,--\,09:35:49 & 
14.6\,+\,44.0 &1374$\pm$40 & 569$\pm$26 & 638$\pm$26 \\
090102.93$+$354928.5 & 3.113 & 47.75 & 31.97 & 2018/04/13\,--\,23:57:23 &
29.4\,+\,84.0 & \pc379$\pm$25 & 128$\pm$14 & 136$\pm$14 \\
090508.88$+$305757.3 & 3.034 & 47.80 & 31.93 & 2017/10/18\,--\,14:20:29 & 
27.8\,+\,71.1 & \pc752$\pm$32 & 307$\pm$20 & 254$\pm$18 \\
094202.04$+$042244.5 & 3.284 & 47.95 & 32.10 & 2017/04/30\,--\,16:47:49 & 
11.2\,+\,54.4 & \pc186$\pm$18 & \pc93$\pm$14 & 160$\pm$15 \\
094554.99$+$230538.7 & 3.265 & 47.25 & 31.78 & 2017/05/08\,--\,11:42:11 &
10.1\,+\,54.3 & $<$\,13 & $<$\,10 & $<$\,16 \\
094734.19$+$142116.9 & 3.034 & 47.81 & 32.12 & 2017/04/28\,--\,16:35:19 & 
24.2\,+\,60.1 & \pc615$\pm$29 & 214$\pm$16 & 230$\pm$17 \\
101447.18$+$430030.1 & 3.122 & 48.16 & 32.38 & 2017/04/16\,--\,03:44:53 & 
21.6\,+\,52.3 & \pc402$\pm$24 & \pc96$\pm$12 & 115$\pm$13 \\
102714.77$+$354317.4 & 3.118 & 48.03 & 32.32 & 2017/10/28\,--\,15:44:35 & 
18.3\,+\,47.6 & \pc713$\pm$30 & 225$\pm$16 & 269$\pm$18 \\
111101.30$-$150518.5 & 3.050 & 47.61 & 31.81 & 2017/05/30\,--\,17:35:32 & 
32.3\,+\,76.2 & \pc163$\pm$23 & \pc41$\pm$10 & \pc47$\pm$11 \\
111120.58$+$243740.8 & 3.193 & 47.77 & 32.09 & 2017/11/17\,--\,23:22:23 & 
24.0\,+\,67.7 & \pc184$\pm$20 & \pc55$\pm$12 & \pc81$\pm$12 \\
114308.88$+$345222.2 & 3.166 & 47.94 & 32.08 & 2017/05/10\,--\,10:22:55 & 
30.3\,+\,76.2 & \pc745$\pm$31 & 240$\pm$18 & 269$\pm$18 \\
114851.46$+$231340.4 & 3.111 & 47.57 & 32.31 & 2017/06/09\,--\,13:34:12 & 
21.8\,+\,43.8 & \pc166$\pm$19 & \pc31$\pm$\pc8 & \pc61$\pm$10 \\
115911.52$+$313427.3 & 3.036 & 47.64 & 32.03 & 2017/05/16\,--\,12:53:01 & 
18.6\,+\,51.5 & \ps27$\pm$14 & $<$\,12 & $<$\,12 \\
120144.36$+$011611.6 & 3.234 & 47.78 & 32.00 & 2017/06/06\,--\,20:38:09 & 
34.8\,+\,82.0 & \pc118$\pm$22 & \pc47$\pm$11 & \pc45$\pm$11 \\
122017.06$+$454941.1 & 3.296 & 47.53 & 31.98 & 2017/11/25\,--\,09:03:59 & 
19.8\,+\,65.9 & \ps64$\pm$15 & \pc29$\pm$11 & \pc12$\pm$\pc9 \\
122518.66$+$483116.3 & 3.096 & 47.62 & 31.96 & 2017/12/21\,--\,07:13:19 & 
24.9\,+\,83.7 & \pc959$\pm$34 & 419$\pm$23 & 461$\pm$23 \\
124637.06$+$262500.2 & 3.114 & 47.73 & 32.01 & 2017/06/27\,--\,14:33:34 & 
32.3\,+\,76.3 & \pc496$\pm$28 & 119$\pm$14 & 150$\pm$15 \\
124640.37$+$111302.9 & 3.155 & 47.63 & 31.82 & 2017/07/03\,--\,16:30:26 & 
32.6\,+\,76.9 & \pc341$\pm$26 & 118$\pm$14 & \pc97$\pm$13  \\
140747.23$+$645419.9 & 3.101 & 47.92 & 32.11 & 2017/11/29\,--\,08:48:35 & 
19.5\,+\,58.4 & \pc420$\pm$25 & 140$\pm$15 & 181$\pm$15 \\
142543.32$+$540619.3 & 3.262 & 47.66 & 32.07 & 2017/11/13\,--\,17:38:58 & 
13.3\,+\,53.4 & \ps22$\pm$13 & $<$\,18 & $<$\,11 \\
142656.18$+$602550.8 & 3.197 & 48.28 & 32.45 & 2017/05/12\,--\,04:55:24 & 
19.8\,+\,47.9 & \pc952$\pm$33 & 236$\pm$17 & 302$\pm$18 \\
145907.19$+$002401.2 & 3.038 & 47.21 & 31.97 & 2017/08/11\,--\,16:24:36 & 
\pc6.5\,+\,39.3 & \ps46$\pm$12 & \pc41$\pm$12 & \pc10$\pm$\pc9 \\
150731.48$+$241910.8 & 3.018 & 46.92 & 31.40 & 2017/07/05\,--\,11:18:06 & 
31.8\,+\,75.3 & \ps45$\pm$12 & $<$\,12 & \pc11$\pm$\pc6  \\
153201.60$+$370002.4 & 3.091 & 47.79 & 32.11 & 2017/06/24\,--\,01:08:49 & 
22.4\,+\,60.6 & \pc208$\pm$21 & \pc77$\pm$12 & \pc64$\pm$11 \\
171227.74$+$575506.9 & 3.001 & 47.77 & 31.99 & 2017/04/29\,--\,03:11:22 &
15.5\,+\,56.6 & \pc646$\pm$30 & 302$\pm$20 & 355$\pm$21 \\
223408.99$+$000001.6 & 3.028 & 47.80 & 32.07 & 2017/05/15\,--\,22:57:49 &
22.2\,+\,60.5 & \pc633$\pm$29 & 173$\pm$15 & 265$\pm$18 \\
\hline
\end{tabular}
\tablefoot{
Columns: (1) source name in the SDSS-DR7 catalogue; (2) redshift from \citet{Hewett-Wild_10} as reported in \citet{Shen+11}; (3) bolometric luminosity from \citet{Shen+11}, in \lumcgs; (4) monochromatic luminosity at rest-frame 2500 \ang computed from a custom fit of the optical spectrum (Lusso et al., in prep.), in \lumcgs Hz$^{-1}$; (5) observation date and time; (6) net exposures for pn and both MOS arrays; (7--9) source net counts in the 0.5--8 keV band, as derived for each EPIC detector from the cleaned event files.
}
\end{table*}

\section{Observations and data reduction}
\label{obs}
The \xmm observations of the 30 quasars in our sample started in April 2017 and were completed over the following year. For each target, the EPIC instruments were operated in Full Frame mode with thin optical filter, with on-source times ranging from 27.9 to 47.6 ks. The observation data files were reprocessed within the Science Analysis System (\sas) v16.1.0, applying the standard filters for background flares. A cut was imposed whenever single-pixel events in the 10--12 keV ($>$10 keV) band exceeded a rate of 0.4--0.6 s$^{-1}$ (0.3--0.4 s$^{-1}$) over the entire pn (MOS) chip, the exact threshold depending on the level of quiescient background. We verified that, in general, this is equivalent to maximizing the signal-to-noise ratio (S/N) in the energy range of interest. We did not resort to any optimised filtering criterion for the faintest objects, as this would artificially boost the detection significance as a consequence of the `Eddington bias', even in the absence of actual background flares. Circular regions with radius of 25$\arcsec$ (pn) and 20$\arcsec$ (MOS) were adopted for the extraction of the source spectra, while the background was evaluated from an adjacent 60$\arcsec$ circle showing no evidence of excess emission. In three cases (J0304$-$00, J0945$+$23, J1507$+$24), the source regions were slightly reduced (to 20$\arcsec$ or 15$\arcsec$ for all detectors) to avoid contamination from a nearby point-like object. Redistribution matrices and ancillary response files were generated with the \sas tasks \rmfgen and \arfgen, respectively. 

By virtue of the suitable quality of these data, we were able to perform a compelling spectral analysis over most of the sample. Indeed, 26 out of 30 quasars at $z\simeq3$ have a total of net EPIC counts in excess of approximately one hundred (namely, 97 for J1459$+$00), and are visually discernible in each image. Only one target (J0945$+$23) turned out to be formally undetected, while the remaining three (J1159$+$31, J1425$+$54, and J1507$+$24) can at least rely on a marginal detection with the pn (Table~\ref{st}). The spectral analysis was carried out with \xspec v12.10.1, rebinning the data to ensure at least one count per energy channel and making use of the $C$-statistic \citep{Cash79,Kaastra17}, as this is more appropriate for the Poissonian (low-count) regime. Unless otherwise stated, the reported uncertainties correspond to a change in the fit statistics of $\Delta C = 1$. For 21 objects, the quality of the pn spectra enabled us to impose a 4$\sigma$ significance per bin, and to subsequently perform a fit with the canonical $\chi^2$ statistic for a consistency check. The results, in terms of both best-fit values and confidence ranges, are always in full agreement with those obtained with the $C$-statistic, on which we thus rely for the remainder of this work. 

For simplicity, we adopt here a standard $\Lambda$CDM cosmology with $H_0=70$ km s$^{-1}$ Mpc$^{-1}$, $\Omega_m=0.3$, and $\Omega_\Lambda=0.7$.

\section{X-ray data analysis}
\label{spe}
\subsection{Hardness ratios}
The X-ray study of high-redshift quasars, even for samples with a limited number of sources, generally relies on hardness ratios (determined by the net counts collected in a soft and a hard band) as diagnostics of the underlying spectral shape. While in this work the data quality allows us to carry out a spectral analysis with sufficient detail, as we show in the following, the use of X-ray colours can still provide some initial hints and support our later findings in a model-independent way. We therefore adopted the fractional difference definition of the hardness ratio, ${\rm HR} = (\mathcal{H}-\mathcal{S})/(\mathcal{H}+\mathcal{S})$, where $\mathcal{H}$ and $\mathcal{S}$ are the source counts in the 0.5--2 keV (soft) and 2--8 keV (hard) bands. As HR also depends on the relative effective area of the detector in the bands of reference, at this stage we considered for simplicity only the pn data. In order to obtain sensible results for all the sources in our sample, including the marginal detections, we made use of the Bayesian Estimation of Hardness Ratios (BEHR; \citealt{Park+06}) code. Indeed, the necessity of resorting to the HR analysis arises in the low-count regime, where the classical Gaussian approximations fail. By avoiding a direct background subtraction, the BEHR method returns reliable errors based on the posterior probability distributions, and it is also applicable in case of non-detection in a given band. 

The estimated hardness ratios and their uncertainties are listed in Table~\ref{rt}. For the brightest objects, there is a strict coincidence with the classical derivation. The median value, $\langle {\rm HR} \rangle = -0.55$, suggests that the spectra are, on average, fairly soft. Taking advantage of the Portable Interactive Multi-Mission Simulator v4.9 (PIMMS)\footnote{\url{http://cxc.harvard.edu/toolkit/pimms.jsp}}, we established that such a ratio, for the mean Galactic column ($\nh^{\rm Gal} \simeq 2.6 \times 10^{20}$ cm$^{-2}$) and source redshift ($z \simeq 3.14$) of the sample, corresponds to a power-law continuum with $\Gamma = 1.79$. Notably, the bulk of the HR distribution is highly symmetric and rather narrow, with a standard deviation of 0.09 when described by a Gaussian centred at ${\rm HR_0} = -0.58$. There are only four outliers towards higher HR values, hence harder spectra: the four marginal detections except J1507+24, plus J1148+23. On these grounds, just a small fraction of objects is expected to require local absorption.\footnote{Throughout the paper, the term `local' refers to the rest frame of a given quasar.} Assuming an intrinsic continuum with $\Gamma = 1.79$ for example, a column density of $\nh(z) = 2.5 \times 10^{23}$ cm$^{-2}$ would be needed to obtain HR = 0.

\begin{figure}
\centering
\includegraphics[width=8.5cm]{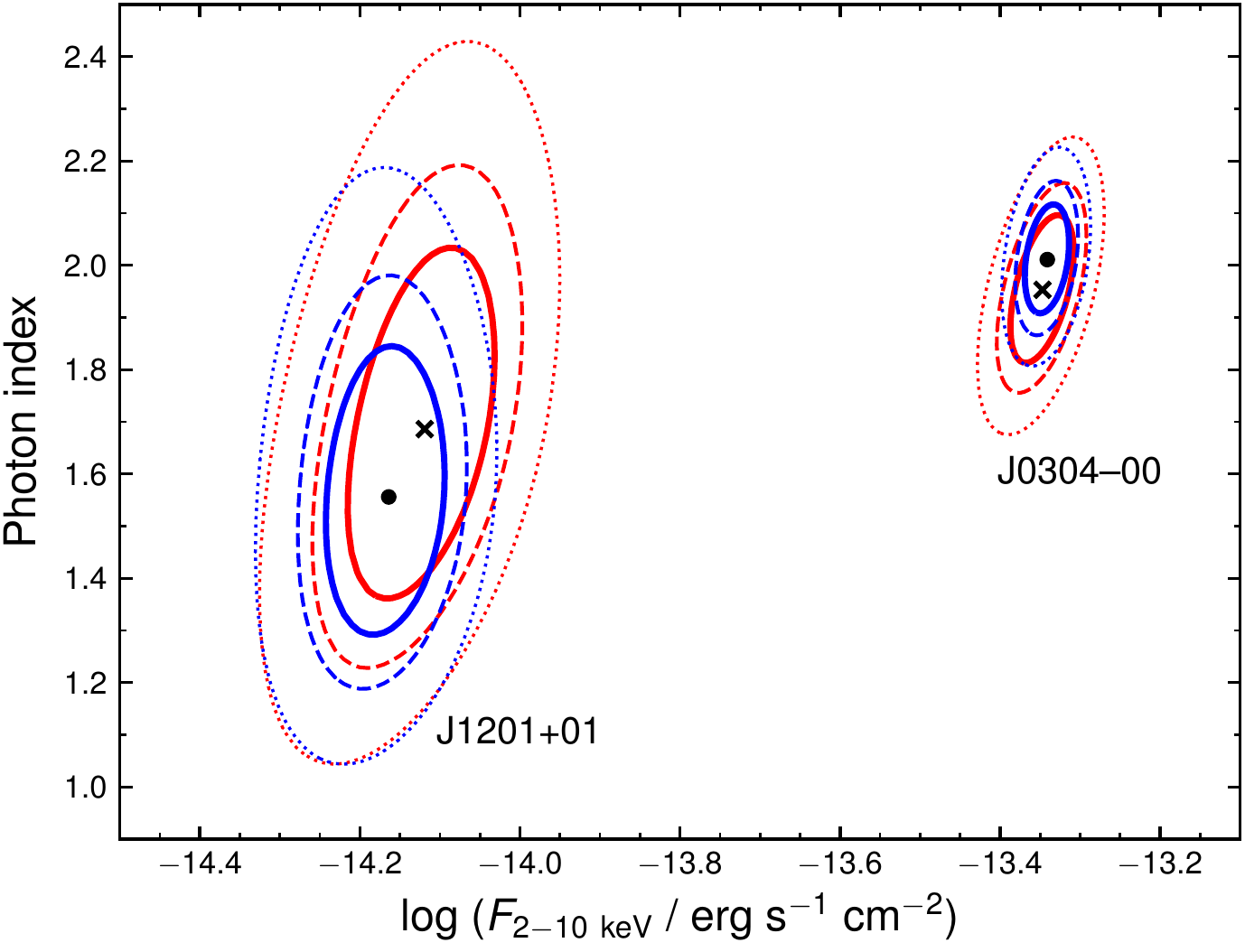}
\caption{Confidence contours corresponding to $\Delta C = 2.30$ (solid), 4.61 (dashed), 9.21 (dotted) in the $\Gamma$ vs. flux plane, as obtained from a separate analysis of pn (blue curves, with best-fit value marked by a dot) and MOS (red curves, cross) spectra. Two examples are shown to be representative of the brightest (J0304$-$00, $\sim$1100 cumulative net counts) and faintest (J1201$+$01, $\sim$200 counts) sources in our sample. The agreement between pn and MOS is always remarkable, and well within the measurement uncertainties even for spectra of relatively low quality.}
\label{tc}
\end{figure}

\begin{figure*}
\centering
\includegraphics[width=18cm]{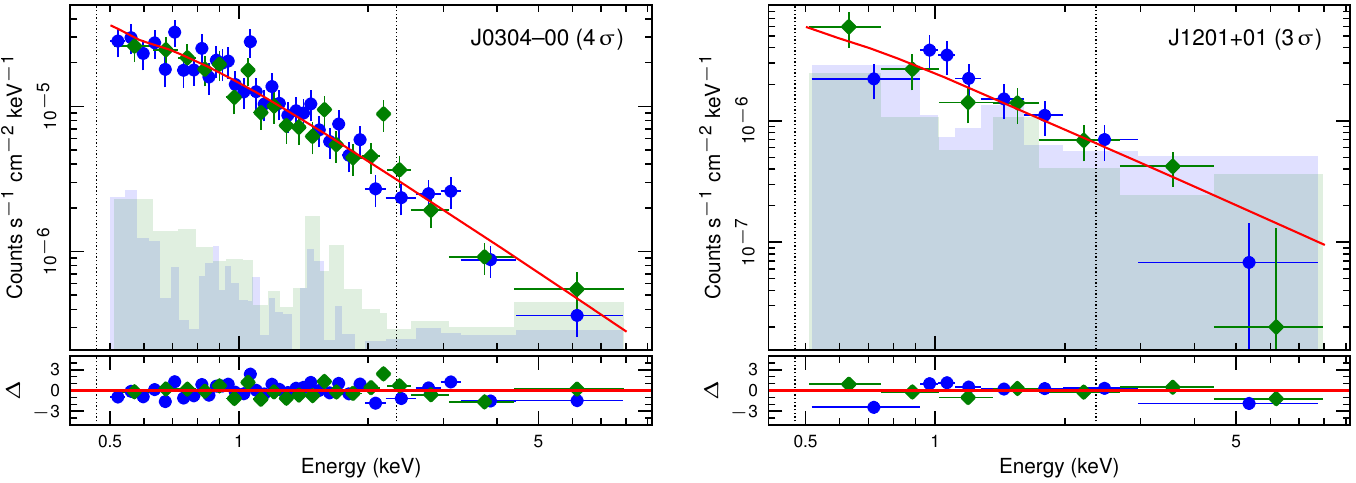}
\caption{Typical \xmm spectra of our $z\simeq3$ quasars, as obtained in this campaign. The same sources already shown in Fig.~\ref{tc}, i.e.~J0304$-$00 (left) and J1201$+$01 (right), have been chosen here to illustrate the full range of spectral quality. Blue dots and green diamonds are used for pn and MOS spectra, respectively, where in the latter case the data from both detectors were merged for visual clarity. The red solid line represents the best-fit power-law continuum with no intrinsic absorption. Residuals with respect to this fit were computed as $\Delta$\,=\,(data$-$model)/error, and are shown in the bottom panels. A graphical rebinning is applied for plotting purposes only so that each energy channel has the significance reported within brackets in the top right-hand corner. The shaded regions indicate the background levels, while the vertical dotted lines mark the rest-frame 2 and 10 keV positions. The spectra of all the other sources, when available, can be found in the Appendix; Fig.~\ref{a1}.}
\label{fs}
\end{figure*}

\begin{table*}
\caption{Results of the X-ray spectral analysis and derived quantities.}
\label{rt}
\centering
\begin{tabular}{cccccccccc}
\hline \hline \noalign{\smallskip}
Source & HR & $\Gamma$ & $C/\nu$ & $\nh(z)$ & $\Delta C$ & $F_{\rm 2-10~keV}$ & log\,$\ldx$ & $\aox$ & $\Delta\aox$ \\
\noalign{\smallskip}
\hline \noalign{\smallskip}
J0303$-$00 & $-0.52^{+0.05}_{-0.06}$ & 1.87$^{+0.08}_{-0.07}$ & 553/608 & $<$\,1.2 & \pn0.0 & \pc3.61$^{+0.17}_{-0.15}$ & 27.59$^{+0.04}_{-0.04}$ & $-$1.68 & $+0.02$ \\
\noalign{\smallskip}
J0304$-$00 & $-0.64^{+0.03}_{-0.04}$ & 1.99$^{+0.05}_{-0.06}$ & 596/660 & 1.1$^{+0.8}_{-0.7}$ & $-$2.5 & \pc4.55$^{+0.15}_{-0.16}$ & 27.75$^{+0.03}_{-0.03}$ & $-$1.64 & $+0.07$ \\
\noalign{\smallskip}
J0826$+$31 & $-0.45^{+0.17}_{-0.15}$ & 1.56$^{+0.17}_{-0.16}$ & 186/213 & $<$\,0.9 & \pn0.0 & \pc1.44$^{+0.15}_{-0.14}$ & 27.03$^{+0.09}_{-0.09}$ & $-$1.87 & $-0.17$ \\
\noalign{\smallskip}
J0835$+$21 & $-0.50^{+0.04}_{-0.05}$ & 1.77$^{+0.07}_{-0.06}$ & 569/676 & $<$\,0.3 & \pn0.0 & \pc3.97$^{+0.14}_{-0.15}$ & 27.55$^{+0.03}_{-0.03}$ & $-$1.67 & $+0.03$ \\
\noalign{\smallskip}
J0900$+$42 & $-0.56^{+0.03}_{-0.02}$ & 1.83$^{+0.03}_{-0.03}$ & 868/1034 & $<$\,0.1 & \pn0.0 & 17.09$^{+0.37}_{-0.37}$ & 28.27$^{+0.01}_{-0.02}$ & $-$1.56 & $+0.20$ \\
\noalign{\smallskip}
J0901$+$35 & $-0.50^{+0.05}_{-0.06}$ & 1.60$^{+0.07}_{-0.08}$ & 527/563 & 0.6$^{+1.0}_{-0.6}$ & $-$0.5 & \pc2.14$^{+0.10}_{-0.10}$ & 27.22$^{+0.04}_{-0.04}$ & $-$1.82 & $-0.12$ \\
\noalign{\smallskip}
J0905$+$30 & $-0.62^{+0.03}_{-0.04}$ & 2.12$^{+0.06}_{-0.05}$ & 668/710 & $<$\,0.3 & \pn0.0 & \pc5.14$^{+0.15}_{-0.15}$ & 27.76$^{+0.03}_{-0.02}$ & $-$1.60 & $+0.11$ \\
\noalign{\smallskip}
J0942$+$04 & $-0.69^{+0.08}_{-0.09}$ & 2.11$^{+0.11}_{-0.10}$ & 385/417 & $<$\,1.6 & \pn0.0 & \pc3.39$^{+0.19}_{-0.18}$ & 27.66$^{+0.05}_{-0.05}$ & $-$1.70 & $+0.02$ \\
\noalign{\smallskip}
J0945$+$23 & $<$\,$-$0.13 & 1.8(f) & 109/123 & 14$^{+20}_{-13}$ & $-$1.4 & $<$\,0.44 & $<$\,26.65 & {\it $-$1.97} & {\it $-$0.30} \\
\noalign{\smallskip}
J0947$+$14 & $-0.64^{+0.04}_{-0.04}$ & 1.88$^{+0.06}_{-0.05}$ & 643/680 & 1.1$^{+0.6}_{-0.6}$ & $-$3.2 & \pc4.90$^{+0.17}_{-0.16}$ & 27.66$^{+0.03}_{-0.03}$ & $-$1.71 & $+0.03$ \\
\noalign{\smallskip}
J1014$+$43 & $-0.58^{+0.05}_{-0.05}$ & 2.21$^{+0.08}_{-0.09}$ & 395/479 & $<$\,0.3 & \pn0.0 & \pc3.19$^{+0.14}_{-0.14}$ & 27.61$^{+0.04}_{-0.03}$ & $-$1.83 & $-0.05$ \\
\noalign{\smallskip}
J1027$+$35 & $-0.60^{+0.04}_{-0.03}$ & 1.91$^{+0.06}_{-0.05}$ & 632/683 & $<$\,0.2 & \pn0.0 & \pc7.08$^{+0.22}_{-0.21}$ & 27.86$^{+0.02}_{-0.03}$ & $-$1.71 & $+0.05$ \\
\noalign{\smallskip}
J1111$-$15 & $-0.59^{+0.13}_{-0.11}$ & 1.71$^{+0.13}_{-0.13}$ & 326/394 & $<$\,0.8 & \pn0.0 & \pc1.00$^{+0.08}_{-0.08}$ & 26.91$^{+0.06}_{-0.07}$ & $-$1.88 & $-0.20$ \\
\noalign{\smallskip}
J1111$+$24 & $-0.49^{+0.09}_{-0.09}$ & 1.77$^{+0.13}_{-0.12}$ & 376/383 & $<$\,0.3 & \pn0.0 & \pc1.33$^{+0.09}_{-0.09}$ & 27.10$^{+0.06}_{-0.06}$ & $-$1.91 & $-0.19$ \\
\noalign{\smallskip}
J1143$+$34 & $-0.58^{+0.04}_{-0.03}$ & 1.94$^{+0.06}_{-0.05}$ & 674/734 & $<$\,0.1 & \pn0.0 & \pc4.52$^{+0.14}_{-0.13}$ & 27.69$^{+0.02}_{-0.03}$ & $-$1.68 & $+0.04$ \\
\noalign{\smallskip}
J1148$+$23 & $-0.25^{+0.10}_{-0.10}$ & 1.16$^{+0.11}_{-0.11}$ & 338/317 & $<$\,1.6 & \pn0.0 & \pc1.21$^{+0.10}_{-0.10}$ & 26.79$^{+0.07}_{-0.07}$ & $-$2.12 & $-0.35$ \\
\noalign{\smallskip}
J1159$+$31 & $-0.08^{+0.51}_{-0.42}$ & 1.8(f) & 186/163 & $<$\,9.3 & \pn0.0 & $<$\,0.34 & $<$\,26.46 & {\it $-$2.14} & {\it $-$0.42} \\
\noalign{\smallskip}
J1201$+$01 & $-0.70^{+0.19}_{-0.14}$ & 1.60$^{+0.14}_{-0.14}$ & 320/348 & 4.0$^{+2.8}_{-2.4}$ & $-$3.0 & \pc0.71$^{+0.06}_{-0.06}$ & 26.78$^{+0.07}_{-0.08}$ & $-$2.00 & $-0.30$ \\
\noalign{\smallskip}
J1220$+$45 & $-0.36^{+0.24}_{-0.19}$ & 1.70$^{+0.31}_{-0.28}$ & 221/240 & $<$\,5.7 & \pn0.0 & \pc0.47$^{+0.07}_{-0.07}$ & 26.66$^{+0.14}_{-0.15}$ & $-$2.04 & $-0.34$ \\
\noalign{\smallskip}
J1225$+$48 & $-0.55^{+0.03}_{-0.03}$ & 1.89$^{+0.05}_{-0.04}$ & 813/882 & $<$\,0.2 & \pn0.0 & \pc6.70$^{+0.17}_{-0.17}$ & 27.82$^{+0.02}_{-0.02}$ & $-$1.59 & $+0.12$ \\
\noalign{\smallskip}
J1246$+$26 & $-0.64^{+0.04}_{-0.05}$ & 2.00$^{+0.07}_{-0.07}$ & 458/565 & 0.9$^{+0.8}_{-0.8}$ & $-$1.2 & \pc2.56$^{+0.11}_{-0.10}$ & 27.44$^{+0.04}_{-0.03}$ & $-$1.75 & $-0.04$ \\
\noalign{\smallskip}
J1246$+$11 & $-0.66^{+0.06}_{-0.06}$ & 2.14$^{+0.10}_{-0.09}$ & 449/504 & $<$\,0.8 & \pn0.0 & \pc1.97$^{+0.10}_{-0.09}$ & 27.39$^{+0.05}_{-0.04}$ & $-$1.70 & $-$0.02 \\
\noalign{\smallskip}
J1407$+$64 & $-0.63^{+0.05}_{-0.05}$ & 2.07$^{+0.08}_{-0.07}$ & 482/555 & $<$\,0.3 & \pn0.0 & \pc3.89$^{+0.15}_{-0.16}$ & 27.65$^{+0.03}_{-0.04}$ & $-$1.71 & $+0.02$ \\
\noalign{\smallskip}
J1425$+$54 & $>$\,0.31 & 1.8(f) & 182/182 & 10$^{+14}_{-7}$ & $-$2.8 & $<$\,0.74 & $<$\,26.88 & {\it $-$1.99} & {\it $-$0.28} \\
\noalign{\smallskip}
J1426$+$60 & $-0.54^{+0.03}_{-0.03}$ & 1.81$^{+0.05}_{-0.04}$ & 685/795 & $<$\,0.5 & \pn0.0 & \pc8.03$^{+0.23}_{-0.22}$ & 27.90$^{+0.02}_{-0.02}$ & $-$1.74 & $+0.04$ \\
\noalign{\smallskip}
J1459$+$00 & $-0.61^{+0.24}_{-0.19}$ & 1.72$^{+0.27}_{-0.24}$ & 199/204 & 3.0$^{+4.5}_{-3.0}$ & $-$0.7 & \pc1.21$^{+0.16}_{-0.15}$ & 26.99$^{+0.12}_{-0.12}$ & $-$1.91 & $-0.20$ \\
\noalign{\smallskip}
J1507$+$24 & $-0.46^{+0.26}_{-0.22}$ & 1.8(f) & 92/121 & $<$\,2.3 & \pn0.0 & $<$\,0.24 & $<$\,26.30 & {\it $-$1.96} & {\it $-$0.34} \\
\noalign{\smallskip}
J1532$+$37 & $-0.54^{+0.08}_{-0.09}$ & 1.69$^{+0.11}_{-0.11}$ & 336/407 & $<$\,1.5 & \pn0.0 & \pc1.59$^{+0.10}_{-0.10}$ & 27.12$^{+0.05}_{-0.06}$ & $-$1.92 & $-0.18$ \\
\noalign{\smallskip}
J1712$+$57 & $-0.46^{+0.04}_{-0.04}$ & 1.68$^{+0.04}_{-0.05}$ & 663/794 & 0.9$^{+0.6}_{-0.6}$ & $-$2.7 & \pc7.54$^{+0.23}_{-0.23}$ & 27.76$^{+0.02}_{-0.03}$ & $-$1.62 & $+0.09$ \\
\noalign{\smallskip}
J2234$+$00 & $-0.52^{+0.04}_{-0.04}$ & 1.86$^{+0.05}_{-0.05}$ & 646/694 & $<$\,0.2 & \pn0.0 & \pc5.49$^{+0.18}_{-0.19}$ & 27.70$^{+0.02}_{-0.03}$ & $-$1.68 & $+0.05$ \\
\noalign{\smallskip}
\hline
\end{tabular}
\tablefoot{
Columns: (1) source ID; (2) hardness ratio relative to the 0.5--2 keV (soft) and 2--8 keV (hard) bands, based on EPIC/pn events only and computed following the Bayesian estimation method of \citet{Park+06}; (3) photon index of the continuum in the unabsorbed (baseline) model, \texttt{phabs}\,$\times$\,\texttt{zpowerlw}, where (f) means that the parameter is frozen; (4) best-fit statistics of the baseline model; (5) column density local to the source in the absorbed model, \texttt{phabs}\,$\times$\,\texttt{zphabs}\,$\times$\,\texttt{zpowerlw}; (6) statistical improvement after allowing for local absorption, whose inclusion is always less significant than the (nominal) 95\% level; (7) intrinsic rest-frame 2--10 keV flux/upper limit as inferred from the baseline/absorbed model, in 10$^{-14}$ \fluxcgs; (8) intrinsic monochromatic luminosity at rest-frame 2 keV, in \lumcgs Hz$^{-1}$; (9) two-point UV (2500 \ang) to X-ray (2 keV) spectral index; (10) deviation of $\aox$ from the $\lx$/$\luv$ relation of \citet{RL19}. Typical statistical uncertainties on $\aox$ and $\daox$ are 0.02 and 0.04, respectively. Entries in italics are upper limits.}
\end{table*}

\subsection{Spectral analysis}
As anticipated, we can obtain a robust measurement of the continuum properties for most objects through X-ray spectroscopy, which was performed over the whole sample. In fact, with the due caveats (see below), we also include the four tentative detections in the analysis. The spectra were fitted in the 0.5--8 keV energy range, below which the relative response of pn and MOS becomes quite erratic, while virtually no source counts are found above 8 keV. In keeping with the blue nature of these objects and with the indications from the preliminary HR analysis, our baseline spectral model simply consists of a power-law continuum only modified by Galactic absorption (from \citealt{Kalberla+05}): in \xspec terminology, this is expressed as \texttt{phabs}\,$\times$\,\texttt{zpowerlw}. We started by fitting the pn and MOS spectra separately, where in the latter case the data from the two detectors were tied together but not combined (we therefore refer to a single MOS spectrum hereafter). There are only two free parameters in the model: the photon index of the power law and the intrinsic rest-frame 2--10 keV flux (assessed through \texttt{cflux} in \xspec, omitted for simplicity from the model definition above). For both quantities, the best-fit values obtained for pn and MOS spectra are fully consistent. The agreement remains largely acceptable also at relatively low S/N. This is demonstrated by Figure~\ref{tc}, where we show the confidence contours in the X-ray continuum slope--intensity plane for two sources, J0304$-$00 and J1201$+$01, respectively falling close to the bright- and faint-end of the sample. Consequently, a joint fit of pn and MOS spectra was performed for each source in the following analysis. 

At $z\sim3$, with the present data quality we would be sensitive to column densities of the order of $\sim$10$^{22}$ cm$^{-2}$. Nevertheless, the unabsorbed model always returns a satisfactory fit with no residual curvature at low energies, suggesting a lack of any obvious photoelectric cutoff apart from that associated to the Galactic foreground column. This clearly emerges from the spectra, best-fit models, and related residuals of J0304$-$00 and J1201$+$01, used again as representative examples in Figure~\ref{fs} (an equivalent plot with the spectra of all the other sources, available in 24 more cases, is provided in the Appendix; Figure~\ref{a1}). Furthermore, the mean (median) observed photon index $\Gamma$ is 1.75 (1.80) over the full sample, and 1.83 (1.86) after neglecting the four tentative detections and the radio-loud source, as expected for the typical continuum slope of radio-quiet type-1 AGNs (\citealt{Piconcelli+05, Bianchi+09}). Unsurprisingly however, the distribution of photon indices (which is discussed in more detail in Section~\ref{dis}) also shows some asymmetry in the form of an extended tail towards lower values, because of the tight correlation between $\Gamma$ and HR. The correspondence between the hardness ratio and observed photon index distributions is very good but not perfect, mainly because the use of MOS data in the spectral analysis introduces some modest rearrangement, while the different Galactic columns have a minor effect. In particular, there are three distinct outliers (J1148+23, J1159+31, and J1425+54) in the range $\Gamma\sim1.0$--1.2 (although uncertainties are very large), which is difficult to reconcile with the standard origin of the X-ray continuum in terms of hot-plasma Comptonization (e.g. \citealt{Lightman-Zdziarski_87, Haardt-Maraschi_93}). 

Notwithstanding the broad success of the baseline model, it is worth checking for the effects of local absorption. The aim is twofold: on the one hand, to understand the nature of the anomalously hard (and weak) X-ray spectra observed in some sources; on the other hand, to obtain an accurate and reliable measurement of the monochromatic flux at 2 keV rest frame, which falls just outside the adopted fitting range and is sensitive to even modest columns. We then allowed for an absorption component at the redshift of the quasar, whereby the modified model now takes the form \texttt{phabs}\,$\times$\,\texttt{zphabs}\,$\times$\,\texttt{zpowerlw}. Here we focus on the good- to high-quality spectra, deferring a customized analysis of the marginal detections to the following section. In 19 out of 26 cases, there is no statistical improvement after the inclusion of the local absorber (Table~\ref{rt}). The $\Delta C = 1$ upper limit on $\nh(z)$, which is the only additional parameter in the fit, has a mean (median) value of 9 (3) $\times 10^{21}$ cm$^{-2}$, and the tightness of the individual constraints shows a clear correlation with the S/N. The other seven sources accept columns of a few $\times 10^{22}$ cm$^{-2}$, but in no case is $\Delta C < -3.84$, equivalent to a 95\% significance in the $\chi^2$ limit. Therefore, the presence of intrinsic X-ray obscuration cannot be firmly established in any of these objects. The remaining four quasars, whose spectral properties can only be loosely determined, merit a separate discussion. 

\subsection{Marginal detections}
With respect to the sources for which only $\approx$15--60 net EPIC counts were collected, we first performed a statistical test of the detection significance, following \citet[see their Appendix A2]{Weisskopf+07}. Specifically, we computed the binomial probability that any excess of counts in the source extraction region is simply due to a positive background fluctuation. The probability of spurious detection, combined over the three detectors, is 0.04 for J0945+23, 0.001 for J1425+54, $3\times10^{-4}$ for J1159+31, and $8\times10^{-11}$ for J1507+24. Notably, these quasars are characterised by the four lowest count rates, and not just shorter net exposures owing to background flares, for example.\footnote{We reiterate that the extraction regions for J0945+23 and J1507+24 are narrower than the default ones, but this has a negligible effect.} This implies that they effectively constitute an unexpectedly faint segment of the sample. While J0945+23 and J1507+24 also have the smallest values of $F_{2500\,\ang}$, this is not enough to fully explain their X-ray weakness (see Section~\ref{dis}). For the latter source at least, the estimated HR falls within the main body of the distribution, suggesting a fairly standard spectral slope. For J1425+54 on the other hand, we only get a lower limit on HR, since most of the pn counts are detected in the hard band. This would correspond to $\Gamma < 0.3$. However, the (few) MOS counts are exclusively soft, implying that in this regime even the hardness ratios reported in Table~\ref{rt} are not entirely dependable.

A preliminary assessment of the monochromatic (2 keV) and integrated (2--10 keV) rest-frame fluxes of these objects was obtained with PIMMS, based on the pn and MOS count rates and assuming $\Gamma = 1.8$. At best, the observed flux densities and fluxes are expected to be of the order of a few $\times 10^{-33}$ \fluxcgs Hz$^{-1}$ and a few $\times 10^{-15}$ \fluxcgs, respectively, which is~well below the rest of the sample. Better constraints are provided by the spectral analysis, which, although basic, is enabled by the use of $C$-statistic despite the low number of counts. With the partial exception of J1425+54, the photon indices from the unabsorbed model are in good agreement with those inferred from the hardness ratios. At the same time, thanks to a pivot effect, the upper limits on the hard-band flux are always within 0.1 dex from the values anticipated with PIMMS. Having confirmed the robustness of the spectral fits also for the faintest sources, as a final step we applied the absorbed model fixing $\Gamma$ to 1.8, in keeping with the average value of the overall distribution. We note that a reasonable range for both $\Gamma$ and $\nh(z)$ cannot be simultaneously determined. Imposing a standard continuum slope modified by a local column leads to a more conservative measure of the intrinsic X-ray intensity. The resulting upper limits are listed in Table~\ref{rt} and are considered from now on for these objects.

\section{Discussion}
\label{dis}

\subsection{Previous X-ray observations}
Among the 30 sources of our $z\simeq3$ \xmm sample, 15 have an archival \chandra snapshot with a duration of between 1.5 and 4.2 ks. In all but one case (J1225+48, which lies $\sim$10$\arcmin$ off-axis in the field of Mrk\,209; ObsID: 10560), these are targeted observations. Due to the short exposures and the smaller effective area of \chandra, the total number of counts at 0.5--8 keV within a radius of 3$\arcsec$ (10$\arcsec$ for J1225+48 to account for the distorted shape of the off-axis point spread function) from the quasar is rather small. Even so, the probability of spurious detection (computed as above, evaluating the background over a circle of 1$\arcmin$ radius) is always of the order of 10$^{-5}$ or much less, except for J1201+01 ($4\times10^{-3}$). Only four objects have enough counts to attempt a spectral analysis. For the other 11, the net counts\footnote{The typical (maximum) background level is of 0.1 (0.2) counts over the source extraction region, and it is therefore neglected.} range from 2 (J1201+01) to 27 (J1027+35), with a median of 4. We can then simply estimate the 2--10 keV flux using PIMMS, with the photon indices of Table~\ref{rt}. In seven sources, the historical X-ray flux is perfectly matched, or consistent within 0.1 dex with the one obtained in the present campaign. The other four objects are compatible with a possible variation in both directions (brightening/fading) by up to a factor of 2.5--3. In principle, luminous quasars harbouring black holes with masses in excess of 10$^9 M_{\odot}$ should not vary significantly in the X-rays on relatively short timescales (e.g.~\citealt{Shemmer+17}).\footnote{The elapsed time in the quasars' rest frame between the \chandra and \xmm observations ranges from about one to four years.} However, there are also examples of a more dramatic behaviour, such as the $z = 6.31$ quasar SDSS\,J1030+0524, whose spectral index flattened by $\Delta\Gamma = -0.6$ with a 2.5 times fainter flux over the (rest-frame) span of only two years \citep{Nanni+18}.  

Better constraints on any possible variability can be derived from the spectra. J0942+04, J1014+35, and J1407+64 were all observed in 2006, for 4.1, 4.2, and 3.8 ks, respectively. Their spectra have been extracted with \ciao v4.11, and have 47($\pm7$), 34($\pm6$), and 47($\pm7$) source counts, respectively. We apply the baseline model and provide in Table~\ref{ht} the inferred power-law photon index and the change of the 2--10 keV flux (in log units) with respect to the entry in Table~\ref{rt}. For completeness, we also analysed the 2006 spectrum of J0900+42 (4.0 ks, 109$\pm$11 net counts). Two more quasars have a previous \xmm observation. J0304$-$00 is serendipitously found in the field of a blazar at $z=0.56$, acquired in Small Window mode so that only MOS data are available for a joint good exposure of 29.8 ks. The total net counts are 100$\pm$11, that is~about one tenth of those collected in 2017. Interestingly, at variance with the more recent spectrum, the slightly brighter and steeper 2004 state does not accept a local absorber, although the $\Delta C = 1$ upper limit of $\nh(z) < 2.3\times 10^{22}$ cm$^{-2}$ is less stringent than the uncertainty range reported in Table~\ref{rt}. J1426+60 had already been observed for 29 ks in 2006, but background flares were quite severe. After the usual filtering, the recovered pn+MOS net exposure is 8.3\,+\,42.5 ks, resulting in 685$\pm$31 counts, only two times less than those employed in our analysis. Consequently, this is the only archival spectrum of sufficient quality for a meaningful comparison. Notably, the power-law photon index was exactly the same as found in 2017, while the intensity has apparently increased by 0.13 dex. Overall, we can therefore conclude that our quasars experience typical variations of $\Delta\Gamma = \pm0.1$--0.2 in slope and $\pm$0.15 dex in flux, even if we cannot rule out larger fluctuations in a few objects.

\begin{table}
\caption{Spectral analysis of archival X-ray observations.}
\label{ht}
\centering
\begin{tabular}{cccc}
\hline \hline \noalign{\smallskip}
Source & Obs. date & $\Gamma$ & $\Delta\log\,(F_{\rm 2-10~keV})$ \\
\noalign{\smallskip}
\hline \noalign{\smallskip}
J0304$-$00 & 2004/07/19 & 2.08$^{+0.22}_{-0.21}$ & $+$0.13$\pm$0.05 \\
\noalign{\smallskip}
J0900$+$42 & 2006/02/09 & 1.97$^{+0.17}_{-0.18}$ & $-$0.18$\pm$0.05 \\
\noalign{\smallskip}
J0942$+$04 & 2006/02/08 & 2.02$^{+0.29}_{-0.27}$ & $+$0.14$\pm$0.07 \\
\noalign{\smallskip}
J1014$+$35 & 2006/06/14 & 1.66$^{+0.33}_{-0.31}$ & $-$0.02$\pm$0.10 \\
\noalign{\smallskip}
J1407$+$64 & 2006/09/16 & 1.76$^{+0.25}_{-0.25}$ & $+$0.10$\pm$0.07 \\
\noalign{\smallskip}
J1426$+$60 & 2006/11/12 & 1.81$^{+0.07}_{-0.07}$ & $-$0.13$\pm$0.02 \\
\noalign{\smallskip}
\hline
\end{tabular}
\end{table}

\subsection{Photon index distribution}
\begin{figure}
\centering
\includegraphics[width=8.5cm]{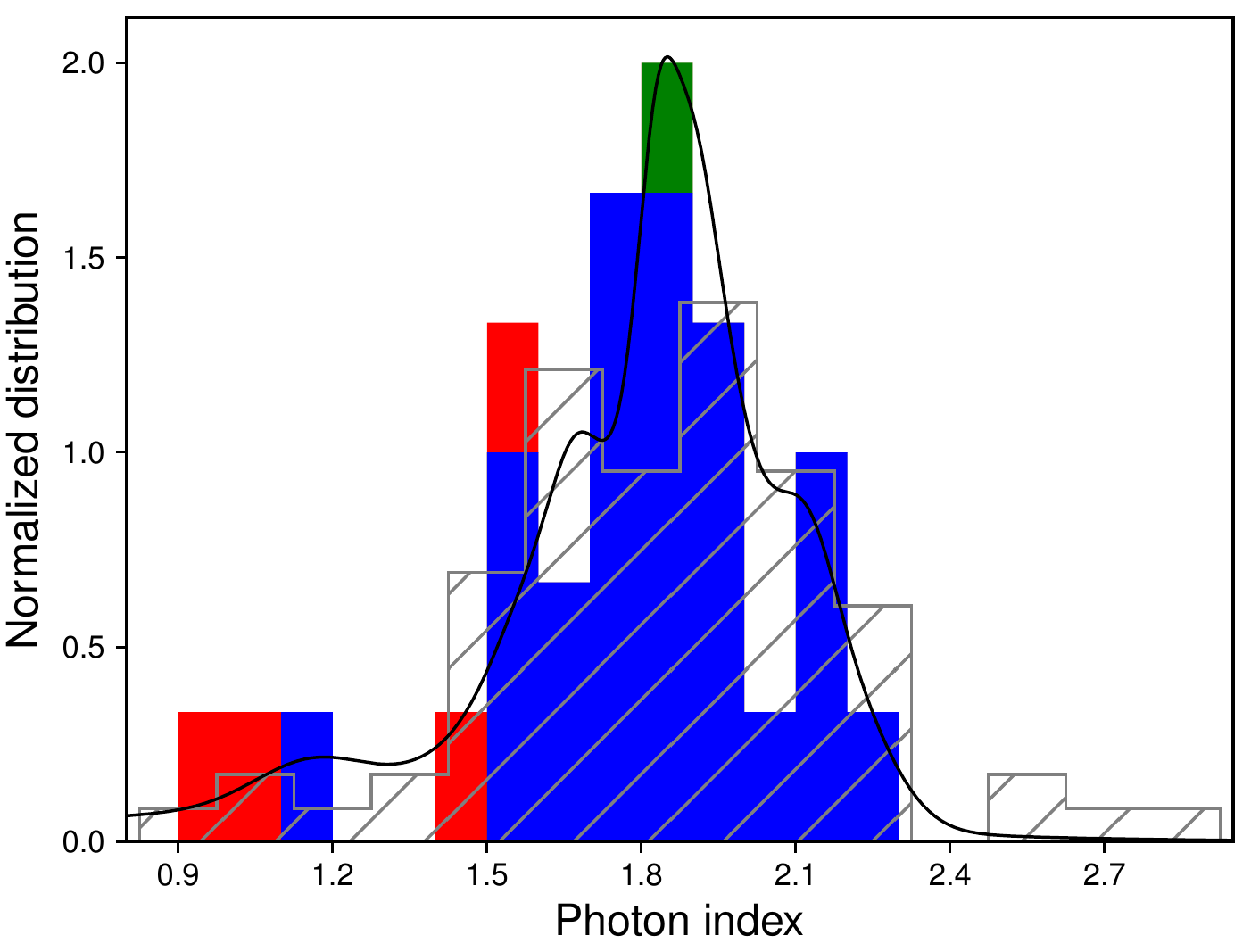}
\caption{Distribution of the power-law photon index in our sample as obtained from the baseline model, \texttt{phabs}\,$\times$\,\texttt{zpowerlw}. The colour code is as follows: green for the radio-loud object, red for the marginal detections, blue for all the other sources. The same convention will be used, when relevant, in all of the following figures. The solid line represents an approximated probability density function that also takes into account the uncertainty in each measurement of $\Gamma$ (see the text for more details). The dotted histogram is the (normalized) distribution of hard X-ray continuum slopes for the quasars in the CAIXA sample, corrected for a luminosity-dependent reflection component \citep{Bianchi+09}.}
\label{dg}
\end{figure}

The size of our sample allows us to draw a statistically informative picture of the X-ray properties of highly luminous, intrinsically blue quasars at $z\simeq3$. As our fits are performed over the $\sim$2--33 keV rest-frame spectral range, which is presumed (also by selection) to be almost unaffected by absorption, we begin with taking into account the shape of the hard X-ray continuum. The normalized distribution of the observed photon index $\Gamma$ is shown in Figure~\ref{dg}, where sources are colour-coded to visually distinguish the radio-loud (1, green) and X-ray faint (4, red) objects from the rest of the sample (25, blue). In principle, this histogram is not necessarily a faithful description of the actual distribution, since about half of the fits give an uncertainty comparable with (or larger than) the adopted bin width, $\Delta\Gamma = 0.1$. Instead of simply considering the best-fitting values of $\Gamma$, we then assumed for each source a normal likelihood distribution, with the average between the upper and lower error bars being the standard deviation. This is a fair approximation, as the uncertainties on $\Gamma$ are largely symmetric (e.g.~Table~\ref{rt}). The composite, re-normalized distribution is also plotted in Figure~\ref{dg}. We note that J0900+42 ($\Gamma = 1.83$) was not removed, since it only contributes to the global amplitude. 

The smoothed probability density function has an even closer similarity, in both centroid and width, to the distribution of the hard X-ray (2--10 keV) power-law slopes found by \citet{Bianchi+09} for the 77 quasars of the CAIXA sample (defined to have an absolute magnitude $M_B < -23$). The only perceptible difference consists in the lack, among our $z\simeq3$ sources, of extremely soft ($\Gamma > 2.4$) X-ray spectra. Such a remarkable match has one major consequence and two corollary ones. First and foremost, this is strong confirmation that the quasar intrinsic X-ray continuum, and therefore the underlying physical mechanism responsible for its origin, does not significantly evolve with redshift or BH mass. In fact, the CAIXA quasars span the redshift range $z \simeq 0.01$--4.52, with only two objects at $z\approx3$, and their median log\,$\mbh$ of 8.3 is about 50 times smaller than that of our sample. Further evidence in this sense has now been gathered up to $z\sim6$ and beyond (see e.g.~\citealt{Vito+19}). In this framework, it would be very unusual if the intrinsic photon index were appreciably different from the observed one (i.e.~steeper). We can therefore rule out a substantial contribution from (i) reflected emission or (ii) local absorption. The former aspect is consistent with the apparent dearth of reprocessing material at high luminosity \citep[and references therein]{Lusso+13}. For the Iwasawa--Taniguchi (or X-ray Baldwin) effect \citep{Iwasawa-Taniguchi_93}, the predicted rest equivalent width (REW) of the fluorescent 6.4-keV \feka feature in our quasars is $\approx$25--40 eV \citep{Bianchi+07}, smaller by a factor of a few with respect to the typical upper limits we can obtain from the spectra by adding an unresolved Gaussian profile over the 6.4--7 keV range. Any luminosity-dependent correction for the associated reflection continuum, under the prescription adopted for the CAIXA sample, would have little effect on the spectral slope ($\Delta\Gamma < 0.05$). In terms of local X-ray absorption on the other hand, the fact that there is no statistical requirement to refine the baseline model would lead to the conclusion that the observed photon index always coincides with the intrinsic one. The extent of the possible deviations is discussed below. 

\begin{figure}
\centering
\includegraphics[width=8.5cm]{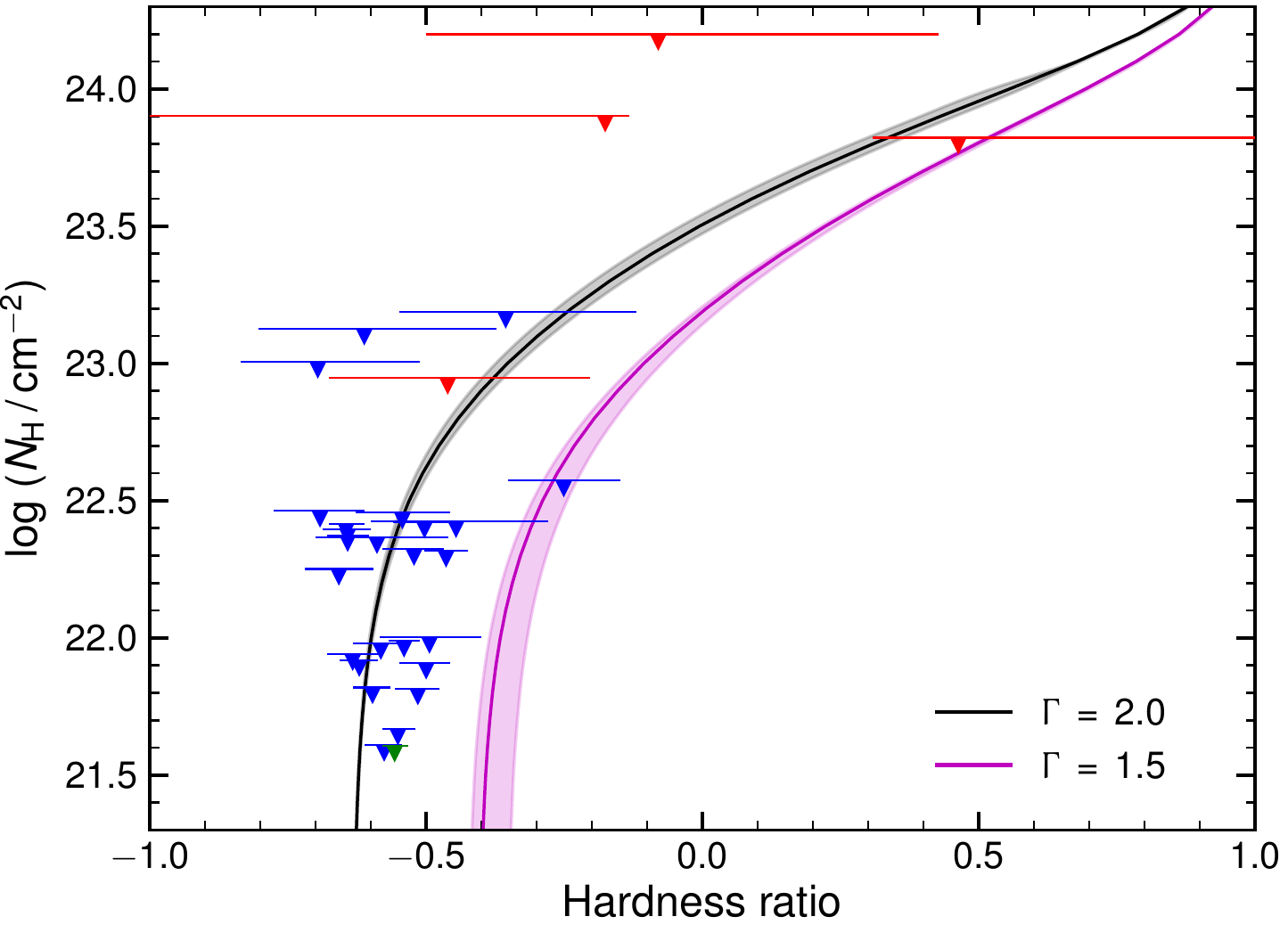}
\caption{Estimated upper limits ($\Delta C = 3.84$, equivalent to the 95\% confidence level in the Gaussian approximation) on the column density $\nh(z)$ as a function of the observed hardness ratio for each source of our sample. Such constraints are obtained from the spectral fits with an absorbed power-law model, where an intrinsic continuum with $\Gamma = 1.8$ was adopted for the faintest objects (red symbols). The synthetic $\nh(z)$--HR curves corresponding to an absorbed spectrum with $\Gamma = 2.0$ (black) and 1.5 (magenta) are also plotted for reference. These have been computed for the mean Galactic column ($\nh^{\rm Gal} = 2.6 \times 10^{20}$ cm$^{-2}$) and quasar redshift ($z = 3.14$), assuming the energy-dependent response of the pn detector. The grey and lilac shaded areas respectively illustrate the uncertainties due to the full range of $\nh^{\rm Gal}$ and $z$ covered by the sample. The large majority of our objects are consistent with being, at most, mildly obscured (i.e. $\nh <$ a few $\times$10$^{22}$ cm$^{-2}$).}
\label{hr}
\end{figure}

\subsection{Constraints on local absorption}
\begin{figure}
\centering
\includegraphics[width=8.5cm]{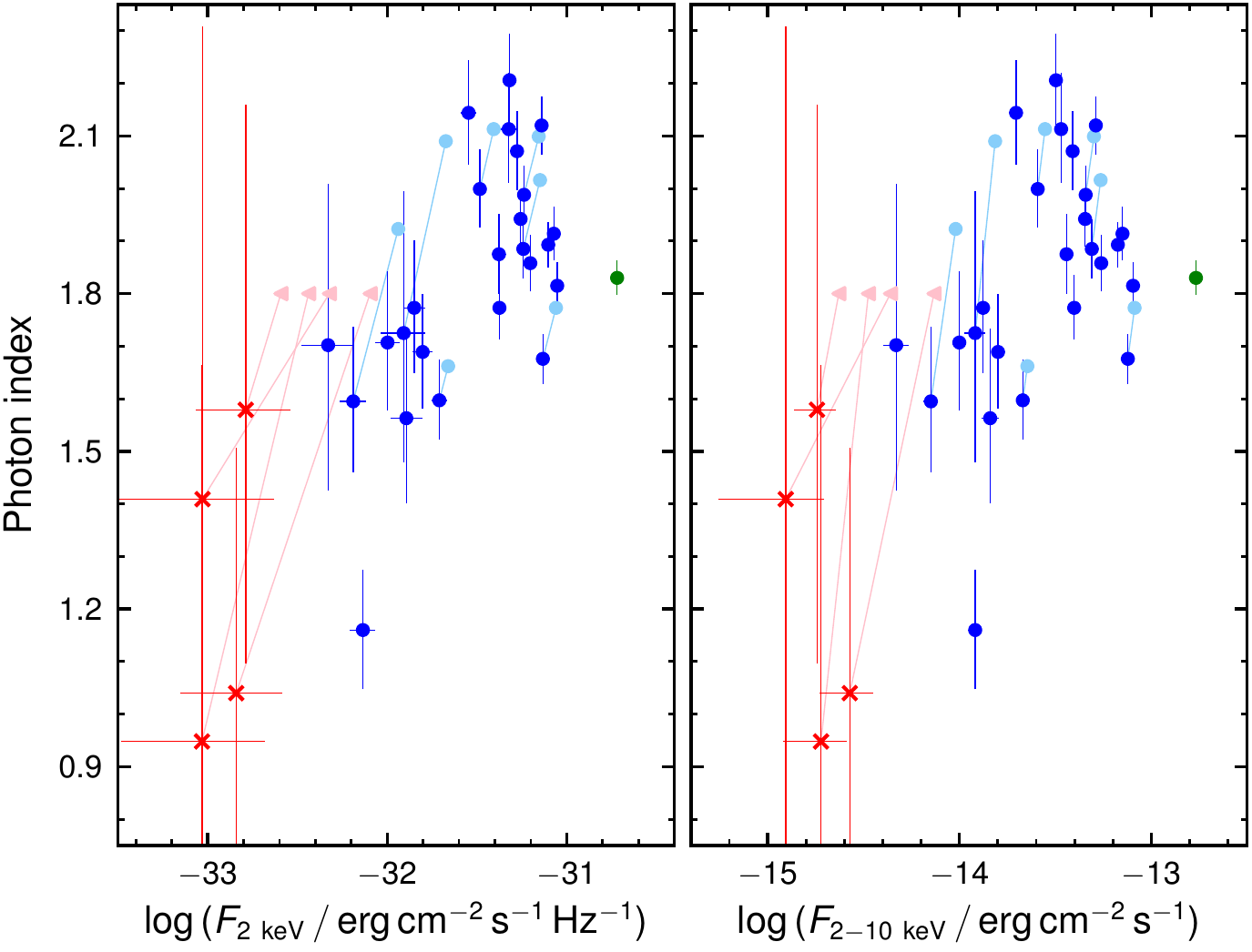}
\caption{Continuum photon index against intrinsic flux density at rest-frame 2 keV (left) and 2--10 keV integrated flux (right). The colour code is the same adopted in the previous figure. The lighter symbols, when present, show the correction required if local absorption were statistically significant. Here the observed best-fit slope is also plotted for the four faintest objects (red crosses), for which an intrinsic $\Gamma = 1.8$ is then conservatively assumed for the sake of discussion (Table~\ref{rt}).}
\label{xp}
\end{figure}

In Figure~\ref{hr}, the $\Delta C = 3.84$ (95\%) upper limits on the column density in the quasars' frame and the corresponding hardness ratios are compared with simulated $\nh(z)$--HR curves, generated with PIMMS for two different values of $\Gamma$. It is immediately evident that any column $\nh(z) < 3\times 10^{22}$ cm$^{-2}$ would have a minor impact on the determination of the photon index, as the observed hardness ratio is only slightly sensitive to a mild obscuration level. This condition is met by the bulk of the sample; indeed, only a handful of objects would accept a column density in excess of 10$^{23}$ cm$^{-2}$. As a more quantitative check, we assume that the non-zero best-fit values of $\nh(z)$ returned by the absorbed model, although not statistically significant, are true. Figure~\ref{xp} shows the continuum photon index against the rest-frame monochromatic (2 keV, left) and integrated (2--10 keV, right) fluxes, corrected for Galactic absorption. When relevant, the shift in the parameter space associated with the putative local absorber is also plotted. Ignoring the faintest subset, in five of the seven sources involved, this is $\Delta \Gamma < 0.1$. The other two (J1201+01 and J1459+00) are the ones with the lowest S/N, meaning that the correction itself, and not just its amplitude, could be ascribed to the poorer data quality. In these two cases, by adopting the baseline (unabsorbed) model, we potentially underestimate the 2-keV flux density by $\sim$75\% and the 2--10 keV flux by $\sim$30\%, respectively, as opposed to the almost negligible 12--24\% and 6--11\% of the other five quasars. Even allowing for these corrections, the main results of the paper would be virtually unaffected. We finally remind that the assumption of $\Gamma = 1.8$ for the marginal detections, which is always steeper than the observed spectral slope (Figure~\ref{xp}), automatically requires a local column $\nh(z)$, and that for this subset we only deal with the upper limits of the absorption-corrected fluxes. 
%

\subsection{X-ray luminosity}
As a whole, our sample is arguably the most X-ray luminous ever observed with regard to radio-quiet quasars, and is definitely one with the highest-quality X-ray spectra to pinpoint our estimates. Restricting ourselves to the core subset of 25 objects, the intrinsic rest-frame 2--10 keV fluxes given in Table~\ref{rt} correspond to luminosities ranging from $4.5\times10^{44}$ to $7.2\times10^{45}$ \lumcgs. For comparison, ULAS J1342+0928, the quasar with the highest known redshift to date ($z=7.54$), has a hard X-ray luminosity of $1.3\times10^{45}$ \lumcgs \citep{Banados+18}. By compiling all the major high-redshift X-ray quasar samples in the literature, only a handful of radio-quiet, non-lensed objects are found at $\log\,(L_{\rm 2-10~keV}/\lumcgs)>45.7$, while with the present analysis we have uncovered four sources above that limit, and as many just below it. Incidentally, the radio-loud J0900+42 reaches out to $\log\,(L_{\rm 2-10~keV}/\lumcgs)\simeq46.2$, and the faintest of our quasars have $L_{\rm 2-10~keV} < 1.2$--$2.5\times10^{44}$ \lumcgs.

\subsection{$\lx$--$\luv$ relation}
The present sample was primarily selected to directly measure the 2-keV flux density with sufficient precision for cosmological applications that rely on a quasar Hubble diagram where luminosity distances are derived from the $\fx$--$\fuv$ relation. As detailed in \citet{RL19}, only 18 out of 29 sources survived the filter on the X-ray photon index. Indeed, for the large majority of their parent sample the observed $\Gamma$ was computed from the soft (0.5--2 keV) and hard (2--12 keV) fluxes reported in the 3XMM-DR7 catalogue \citep{Rosen+16}, using 1.05 and 3.1 keV as pivot points based on the energy-dependence of the EPIC effective area.\footnote{We refer the interested reader to the online Supplementary Material of \citet{RL19} for a complete description of this procedure.} A conservative cut on this `photometric' photon index at $\Gamma > 1.7$ was then adopted to minimize the contamination from absorbed objects. 

\begin{figure}
\centering
\includegraphics[width=8.5cm]{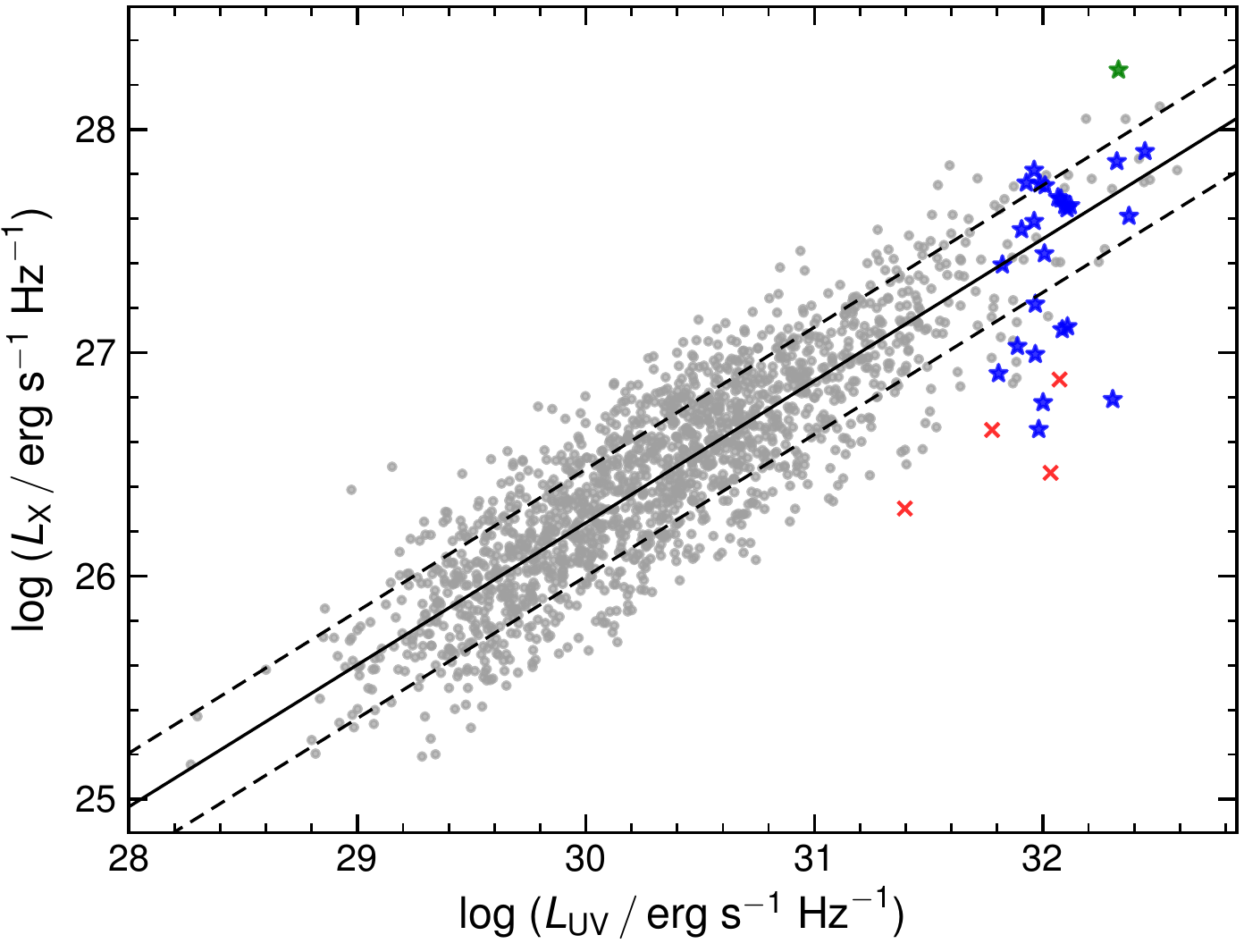}
\caption{Rest-frame monochromatic luminosities $\lx$ against $\luv$ for the 30 quasars in our \xmm $z\simeq3$ sample, with the same colour code as the previous figures (red crosses indicate the upper limits in $\lx$ for the faintest objects). The grey dots represent the sample of about 1600 quasars from \citet{RL19}, with the relative regression line. The dashed lines trace the 1$\sigma$ dispersion, 0.24 dex.}
\label{lr}
\end{figure}

When applied to our quasars, which have targeted observations and good-quality spectra, such a criterion is definitely too crude. Despite this, the sample is far less uniform in the X-rays than it is in the UV, where the 1$\sigma$ dispersion on $L_{2500\,\ang}$ is only 0.1 dex (Table~\ref{st}). In particular, our accurate spectral analysis brings out a clear correlation between $\Gamma$ and flux (Figure~\ref{xp}), whereby fainter sources also display a flatter X-ray continuum. We therefore expect that most of the objects rejected as unsuitable cosmological probes because of $\Gamma$ will strongly depart from the $\lx$--$\luv$ relation. This is confirmed by Figure~\ref{lr}, where all 30 sources are superimposed on the $\lx$--$\luv$ relation obtained by \citet{RL19} for their clean final sample of approximately 1600 quasars. The simple visual inspection reveals that a good fraction of our $z\simeq3$ quasars fall considerably (by factors of $\approx$5) below the correlation, whose scatter of 0.24 dex is essentially constant over at least three orders of magnitude in $\luv$. As discussed above, this behaviour cannot simply be ascribed to obscuration. Another possibility is that for some (as yet unknown) reason the X-ray corona is undergoing a radiatively inefficient phase in about one-third of our objects.

\subsection{X-ray weakness fraction}
\begin{figure}
\centering
\includegraphics[width=8.5cm]{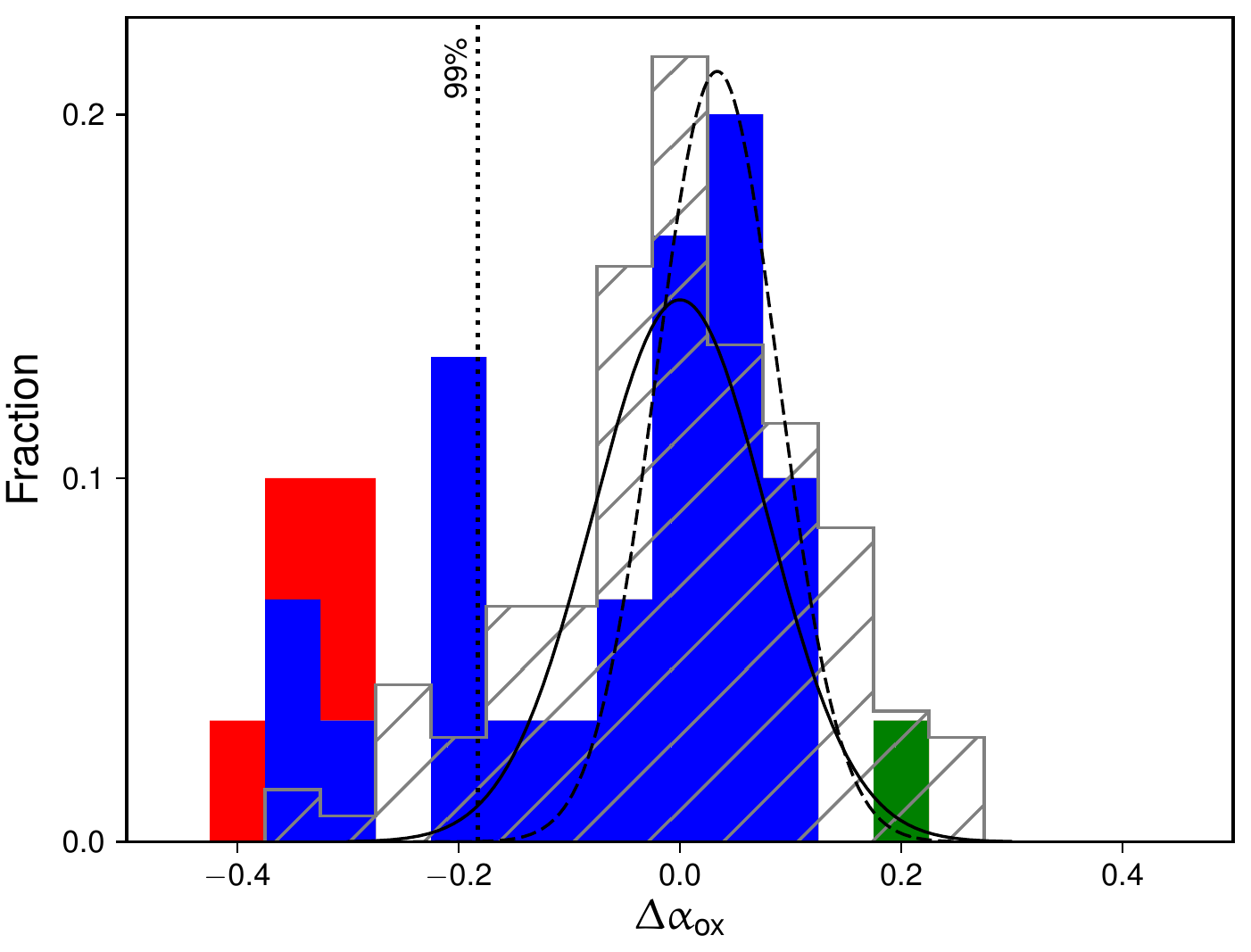}
\caption{Distribution of the differences between the observed $\aox$ and the value predicted from the $\lx$--$\luv$ correlation for all the quasars in the $z\simeq3$ sample. The dashed and solid curves are the best-fitting normal distributions for the core subset of the 25 radio-quiet sources (plotted in blue), where the peak position is left free to vary or forced to be at $\daox=0$. The dotted vertical line marks the one-sided 99\% probability for the latter curve. Irrespective of the exact assumptions (see Section\,\ref{dis}), a substantial fraction of objects clearly fall in the `X-ray weak' tail. The hatched grey distribution refers to `sample B' of \citet{Gibson+08}.}
\label{hd}
\end{figure}

In order to define a quantitative criterion of `X-ray weakness',\footnote{Since the first identification with \rosat (operating at 0.1--2 keV) of quasars with reduced soft X-ray emission compared to their optical flux (e.g.~\citealt{laor+97}), the term `X-ray weak' has been widely used irrespective of the actual origin of this deficit. Conversely, here we favour by default an intrinsic weakness, supported (with some declared caveats) by the outcome of the spectral analysis.} we first fit over the redshift range covered by our quasars a relation of the form (log\,$\fx$+31.5)\,=\,$\gamma$(log\,$\fuv$+27.7)+$\beta$, where fluxes are cosmology-independent. By selecting only the other 30 objects found at $z=3.0$--3.3 in the clean sample of \citet{RL19}, we get a slope $\gamma=0.564\pm0.088$ and an intercept $\beta=-0.326\pm0.045$, with a dispersion of 0.21 dex. From these parameters, we then compute for a given $\fuv$ the expected slope of the power law connecting the $(\nu,F_{\nu})$ points at rest-frame 2500\,\ang and 2 keV in a quasar's SED, $\aox = 0.384\,\log\,(F_{2\,{\rm keV}}/F_{2500\,\ang})$. The non-linearity of the X-ray to UV correlation implies that the SED becomes steeper with higher UV luminosities, thereby inducing the well-known anti-correlation between $\aox$ and $\luv$ \citep[and references therein]{LR17}, which serves as the natural benchmark to determine the extent of any `intrinsic' X-ray weakness. The distribution of the differences $(\daox)$ between the observed and predicted values of $\aox$ is plotted in Figure~\ref{hd}, with the usual colour code and a bin size of 0.05, comparable with the maximum statistical uncertainty on $\daox$. A sizeable fraction (about one-third) of the sample lies relatively far from the centre of the distribution, to the point that a hint of a secondary detached peak emerges at $\daox \la -0.2$, symmetric to the position of the radio-loud source, which is expected to be a clear outlier in the opposite direction. 

Neglecting J0900+42 and the four marginal detections, we model the $\daox$ distribution with a Gaussian shape. The central value, when left free to vary, is within a bin width from the best-fit $\aox$--$\fuv$ relation at $z=3.0$--3.3, at $\daox \simeq 0.04$, while the standard deviation is 0.05. Forcing instead the peak to be at $\daox = 0$ we obtain an equally good description, but with a more conservative dispersion of 0.08, which will be used hereafter to define the statistical threshold for X-ray weakness (note that doubling the bin size actually returns a slightly smaller value, 0.07). Both curves are shown in Figure~\ref{hd}, respectively as the dashed and the solid black lines. Assuming the one-sided 99\% probability, we have 11 (seven plus four) objects that can be considered as X-ray weak. Conversely, only J0900+42 is X-ray loud. Depending on the exact threshold, and also taking into account the possible uncertainty (both statistical and systematic) on $\daox$, the X-ray weakness fraction in our sample thus ranges from 24\% (7/29, excluding the four quasars at $\daox \approx -0.2$ that are broadly consistent with the wing of the main distribution; see also Figure~\ref{lr}) to 38\% (11/29). Taken at face value, the four marginal detections would represent a hard lower limit of 14\%. Our provisional best guess is therefore of $\approx$25\%.

\subsection{Comparison with other samples}

Even in the most conservative scenario, the fraction of X-ray weak quasars in our sample is not only surprising, but also much larger than suggested by previous works. The WISSH quasars analysed in the X-rays, for instance, perfectly follow as a whole the extrapolation of the $\lx$--$\luv$ relation for lower luminosity objects (\citealt{Martocchia+17}; see their Figure~5). Therefore, the reported prevalence of low X-ray-to-optical flux ratios is largely a by-product of the $\lx$--$\luv$ relation itself, rather than a sign of genuine X-ray weakness within the WISSH sample. 

Other notable studies have typically explored lower redshifts and bolometric luminosities. A straightforward comparison can be made with the so-called `sample B' from \citet{Gibson+08}, who analysed 536 SDSS quasars at $z=1.7$--2.7 with archival X-ray data. Their sample B contains 139 radio-quiet, non-BAL quasars observed with \chandra for at least 2.5 ks and lying less than 10$\arcmin$ off-axis, all of which are detected. A normal fit to the $\daox$ distribution, self-consistently computed from the $\aox$--$\luv$ correlation, gives a standard deviation of $\approx$0.09, in very good agreement with our results. The minimum $\daox$ is $-0.37$, suggesting that less than 2\% of optically selected quasars are X-ray underluminous by a factor of ten or more. However, the $\daox$ distribution for sample B is much more symmetric than ours, with a gentle, smoother wing at negative values (Figure~\ref{hd}). By adopting the same one-sided 99\% probability threshold as above, adjusted to the dispersion in $\daox$ of sample B, the objects that qualify as X-ray weak are just about 8\%. The probability that our 29 radio-quiet quasars and sample B are drawn from a single parent distribution was assessed through a Kolmogorov--Smirnov test, treating all of our data as uncensored, and amounts to 1.1\%.

\begin{figure}
\centering
\includegraphics[width=8.5cm]{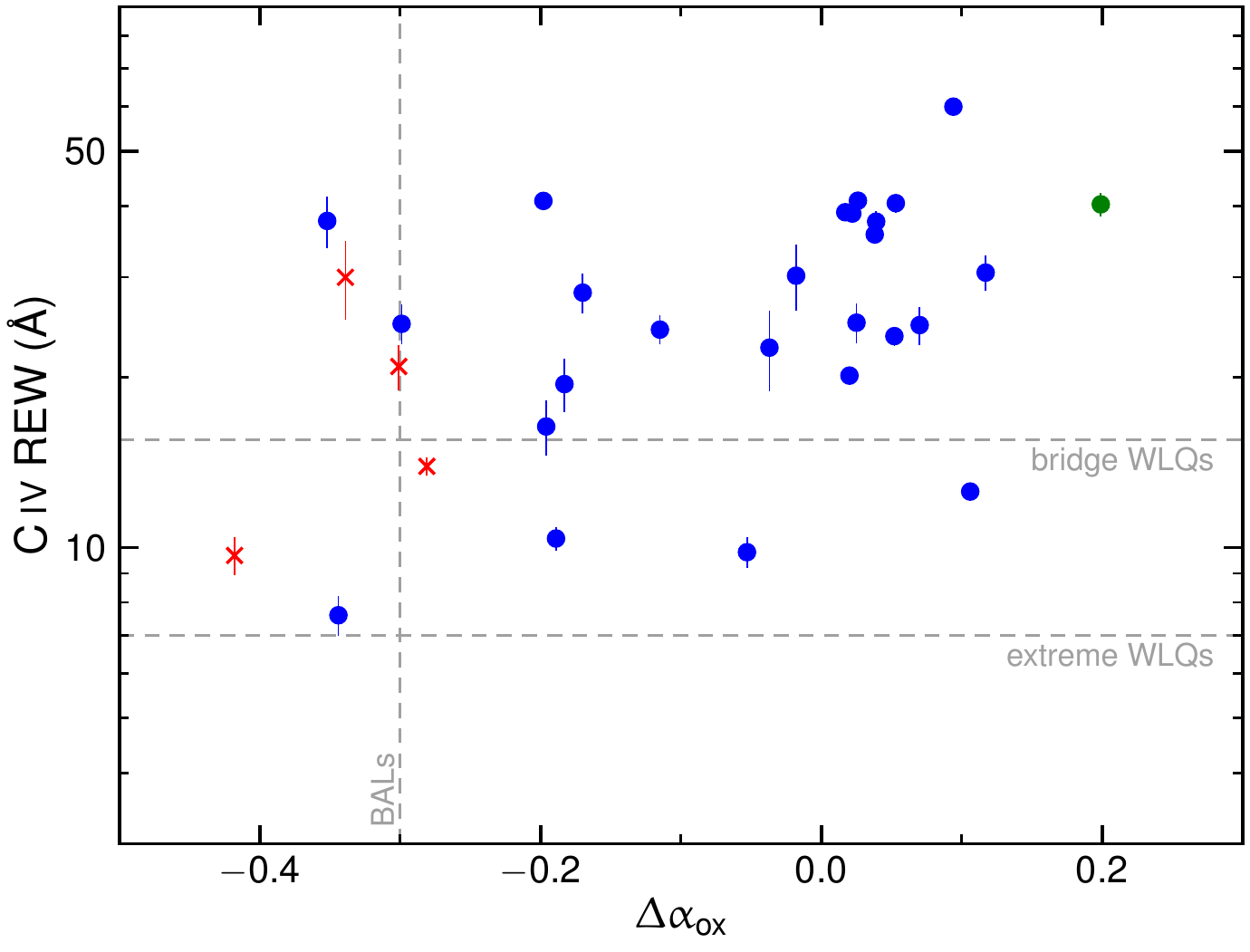}
\caption{Rest equivalent width of the \civ $\lambda$1549\ang emission line, as obtained from our custom fits of the SDSS spectra, plotted against $\daox$. There is a hint of a possible correlation, confirmed at the 98.5\% significance by the Pearson's $\rho$ and Kendall's $\tau$ tests. For reference, we also show the typical thresholds for two quasar populations known to exhibit enhanced rates of X-ray weakness, that is,~BAL ($\daox < -0.3$; \citealt{Liu+18}) and weak-line (extreme subset: \civ REW\,$<$\,7~\ang; bridge subset: \civ REW $=7$--15.5~\ang; \citealt{Ni+18}) quasars.} 
\label{ew}
\end{figure}

A larger fraction of intrinsically X-ray weak quasars is usually found in the BAL population ($\approx$6--23\%; \citealt{Liu+18}). The X-ray weakness of BALs has been traditionally attributed to obscuration, as some kind of shielding of the outflowing gas is required to prevent over-ionization and facilitate line-driving (e.g.~\citealt{Murray+95}). Evidence is however growing in support of the idea that many BALs might actually be intrinsically X-ray weak \citep{Teng+14,Luo+14}, that is,~emitting much less in the X-rays than dictated by the $\lx$--$\luv$ relation. Weak-line quasars constitute another class that would enhance the incidence of X-ray weak sources in our sample. Indeed, while excluding BALs, none of our selection criteria are based on the emission-line properties. Different definitions have been used in the literature for weak-line objects. Here, we follow the convention of \citet{Ni+18}, who distinguished between an extreme subsample with \civ$\lambda$1549\ang REW\,$<$\,7~\ang and a bridge subsample with \civ$\lambda$1549\ang REW $=7$--15.5~\ang. The interpretation for weak-line quasars has been adapted from that of BALs, where in this case it is the geometrically thick inner disc at high accretion rates that shields the broad-line region from ionizing photons. Up to one half of weak-line quasars is also X-ray weak \citep[and references therein]{Luo+15}, and at least some are intrinsically so \citep{Leighly+07}. Within the uncertainties on the equivalent width estimated from our custom fit of the SDSS spectrum, seven quasars in our sample have a moderately weak \civ emission line. Five of them are X-ray weak (Figure~\ref{ew}). As already realized by \citet{Gibson+08}, there is a possible correlation between $\daox$ and log\,(\civ REW), significant at the $>$98.5\% level according to both Pearson's $\rho$ and Kendall's $\tau$ tests. The discussion of this effect goes beyond the scope of this study, and is deferred to a future paper (Lusso et al., in prep.). 

\subsection{Origin of X-ray weakness}

In summary, the explanation of the anomalous X-ray weakness fraction in our $z\simeq3$ sample might not be univocal, as the quasars involved likely represent a mixed bag of objects. Some more clues can come from a one-by-one examination. We start with the four X-ray weak candidates found at $\daox \approx -0.2$ (Table~\ref{rt}, Figure~\ref{hd}). With the tentative correction for local absorption, J1459+00 would move straight into the X-ray normal population, thanks to an increase by 0.09 of its $\aox$. Unfortunately, this \xmm observation was plagued by background flares, meaning that the spectral quality is not enough to ascertain the role, if any, of X-ray obscuration. A negative fluctuation must instead be presumed for J1532+37, yet compared to the 2012 \chandra snapshot, there is no evidence that the source was caught in a fainter-than-usual state. The same holds for J1111+24, which however falls in the weak-line quasar class alongside J1111$-$15. Among the remaining seven objects with $\daox \approx -0.3$, J1159+31, J1220+45 and J1425+54  also have a weak \civ line. This could account in itself for their X-ray weakness. We note that the X-ray weakness fraction corresponding to the range in \civ REW of 7--15.5~\ang is marginally larger in our sample than in \citet{Ni+18}, that is~5 out of 7 versus 7 out of 16. In the latest SDSS data release (DR14; \citealt{Paris+18}), J0945+23 and J1148+23 have been flagged as \civ BALs, previously unidentified in the \citet{Shen+11} DR7 catalogue. While J0945+23 has an  otherwise blue continuum, we suspect a more complex BAL system in J1148+23. Curiously, its X-ray spectrum is abnormally flat ($\Gamma \sim 1.2$), but it does not accept any local absorber. With $\Gamma$ fixed to 1.8, the fit deteriorates by $\Delta C = 12$. Finally, for the last two sources we must allow for both X-ray obscuration and flux variability, yet the former is ruled out in J1507+24 and the latter in J1201+01, based on the \xmm and \chandra observations, respectively.  

Even considering all the systematics above, it is not obvious that the X-ray weakness fraction of our blue quasar sample can be reconciled with the one of sample B from \citet{Gibson+08}. The contamination from any missed BALs is indeed a critical issue overall (see also Appendix~\ref{apb}), but it is restricted to a few objects at most, while the correction for $\nh(z)$ is generally minor. X-ray variability should not cause a net shift in the $\daox$ distribution. A strong bias in favour of negative fluctuations, perhaps tolerated by the limited statistical size of our sample, is at odds with the fact that the distribution peaks at positive values. Finally, weak-line quasars are
also included with a commensurate percentage (although slightly smaller) in sample B, and they are not filtered out. Dismissing the possibility that these effects concur in the same direction as overly fine-tuned, we are left with the necessity for a physical justification for the X-ray weakness of a luminous blue quasar. Any such mechanism seemingly leads to a different state of the X-ray-emitting region, whereas the properties of the UV disc are virtually unchanged.

In common with BAL and weak-line quasars, our sources boast an accretion rate extending across the Eddington limit, with an average value (without J0900+42) of 0.92 assuming the virial BH mass from \civ in \citet{Shen+11}.\footnote{We are currently acquiring near-infrared spectra in the rest-frame \hb region to achieve a more accurate measure of the BH masses.} In this regime, the extreme physical conditions are conducive to the launch of powerful winds, which can drive away most of the mass accreted through the outer disc (e.g.~\citealt{Nardini+15}). In the presence of a disc wind, even without invoking any shielding, the X-ray corona might be intrinsically starving if a significant fraction of the gravitational energy is not actually converted into optical and UV radiation (i.e.~the seed photons for Compton up-scattering) but is instead dissipated to provide the wind with the required thrust. A differential mass accretion rate across the outer and the inner disc (as suggested, albeit inconclusively, for the sample of \citealt{Capellupo+16}), would have a modest impact on the colder portion of the SED, but dramatic changes could emerge in the extreme-UV and X-ray domains \citep{Slone-Netzer_12,Laor-Davis_14}. To first approximation, a highly negative $\daox$ value might be itself an indicator of the amount of gravitational energy lost through an accretion-disc wind. 

Besides the proposed `coronal starving' scenario, there are other viable channels through which an (even failed) inner-disc wind could suppress the observed X-ray emission. Any clump of highly ionized gas in front of the X-ray source, for instance, would scatter a substantial number of X-ray photons out of the line of sight. As a variation on the theme of shielding however, this entails a relatively small covering factor but no intrinsic weakness. A direct quenching is possible instead, if the dense, {\it cold} wind becomes intermingled with the tenuous, hot corona, thus hindering magnetic buoyancy and/or promoting bremsstrahlung rather than inverse Compton cooling \citep{Proga_05}. This is mostly effective in a radially extended corona above the disc, which might yet cause some tension with the X-ray compactness inferred from microlensing (e.g. \citealt{Mosquera+13}).

All these conjectures call for an in-depth broadband analysis, and will be further investigated in the subsequent papers of this series. It is however clear that fundamental insights into the physics of the $\lx$--$\luv$ relation can also be gathered from the outliers. Establishing whether the broadband SED shape retains the signature of winds would revolutionise our way of assessing the role of radiative feedback at the peak of the quasar epoch before the advent of the next-generation X-ray observatories.

\section{Conclusions}
\label{con}
Here we present the X-ray analysis of 30 quasars at $z \simeq 3.0$--3.3, observed as a part of an \xmm Large Programme in 2017--2018 and selected in the optical from the SDSS-DR7 to be representative of the most luminous, intrinsically blue quasars at high redshift. This is a unique sample, put together to further test the suitability and effectiveness of quasars as cosmological standard candles and so benefitting from an unprecedented degree of uniformity. Our main results can be summarised as follows: 
\begin{itemize}
\item Excluding the radio-loud quasar, for 25 out of 29 sources we were able to perform a proper spectral analysis, thanks to the availability of a few to several hundred net counts. The rest-frame 2--10 keV fluxes are in the range 0.5--$8 \times 10^{-14}$ \fluxcgs, which correspond to luminosities of $\log\,(L_{\rm 2-10~keV}/\lumcgs) \simeq 44.6$--45.9. 
\item Four sources turned out to be very faint, but only one is formally undetected, at a spurious detection level of 4\%. 
\item The probability density function derived from the observed photon index distribution peaks at $\Gamma \simeq 1.85$, and its overall shape is in excellent agreement with those obtained in the literature for quasars of lower redshift, luminosity, and BH mass. This corroborates the notion that the physical mechanism responsible for the intrinsic X-ray emission of quasars does not evolve with cosmic time and is scale-invariant. 
\item X-ray absorption in the source frame is never statistically required by the spectral fits. In most objects, a local column in excess of $\nh(z) > 3\times 10^{22}$ cm$^{-2}$ can be safely ruled out. 
\item Based on the archival X-ray data (mostly consisting of very short snapshots) of 17 sources, our quasars show a typical flux variability of $\pm$0.15 dex over a few years, as usually observed in high-redshift quasars with similar BH masses.  
\item Despite the UV homogeneity of the whole sample, the comparison with the $\lx$--$\luv$ relation reveals two rather distinct X-ray populations. About two-thirds of our quasars cluster around the relation, with a minimal dispersion of 0.1 dex. The remaining one-third appear to be moderately to significantly X-ray underluminous, by factors of $>$\,3--10.
\item The X-ray weakness fraction among our $z\simeq3$ blue quasars ($\approx$25\%) is undoubtedly larger than previously reported for radio-quiet, non-BAL quasars at lower redshift and luminosity. While this is likely a miscellaneous subset, we speculate that in some cases the X-ray corona might be in a radiatively inefficient state for the presence of an accretion-disc wind.
\end{itemize}

By construction, this quasar sample stands out as ideal for understanding how the captivating $\lx$--$\luv$ correlation is rooted in the workings of SMBH accretion. In this context, outliers could denote a glitch in the transfer of the gravitational energy of the infalling matter to the X-ray corona. In our subsequent papers we will focus on the multiwavelength properties of each source, to determine if and how these reflect the discovered X-ray dichotomy. 

\begin{acknowledgements}
We thank the anonymous referee for their useful comments and suggestions. This research was finalized at the Aspen Center for Physics (ACP), which is supported by National Science Foundation grant PHY-1607611. We acknowledge financial contribution from the agreement ASI--INAF n.\,2017-14-H.O. EN acknowledges funding from the European Union's Horizon 2020 research and innovation programme under the Marie Sk\l{}odowska-Curie grant agreement no. 664931, and partial support from the Simons Foundation during the ACP stay. The results presented in this paper are based on data obtained with \xmm, an ESA science mission with instruments and contributions directly funded by ESA member states and NASA. For catalogue cross analysis we have used the Virtual Observatory software \topcat \citep{Taylor_05}, available online at \url{http://www.star.bris.ac.uk/~mbt/topcat/}. The figures were generated with \matplotlib \citep{Hunter_07}, a \python library for publication-quality graphics. 
\end{acknowledgements}

\begin{appendix}

\section{Gallery of X-ray spectra}

We present in Figure~\ref{a1} the pn (blue dots) and MOS (green diamonds, merged) spectra of 24 sources in our sample. Those of J0304$-$00 and J1201+01 have already been shown in Figure~\ref{fs}, while in the other four cases the number of net counts was too low to obtain a detailed spectrum. For completeness, in each panel we also plot the best-fit baseline model (with no local absorption), the relative residuals, and the background levels.

\begin{figure*}
\centering
\includegraphics[width=18cm]{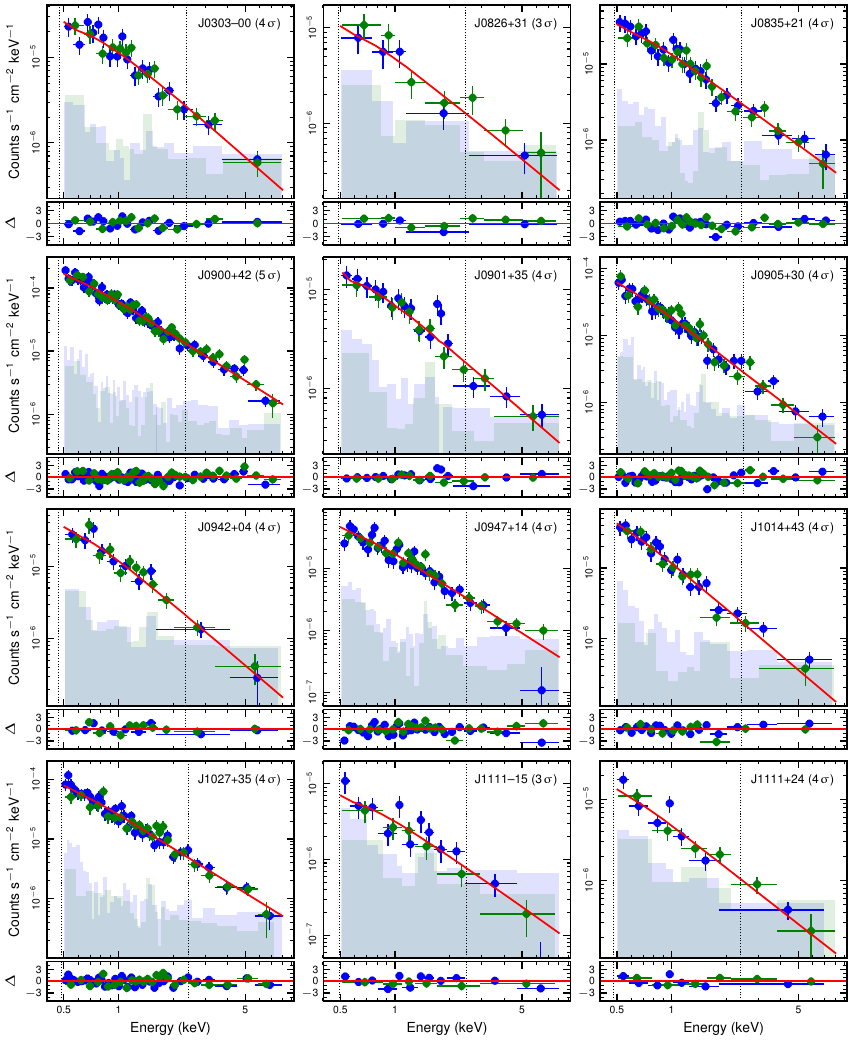}
\caption{\xmm spectra of the sources in our $z\simeq3$ sample, rebinned for graphic purposes only to the statistical significance indicated within brackets. The complete legend is the same as in Figure~\ref{fs}, with the vertical lines marking the rest-frame energies of 2 and 10 keV.}
\label{a1}
\end{figure*}

\begin{figure*}
\addtocounter{figure}{-1}
\centering
\includegraphics[width=18cm]{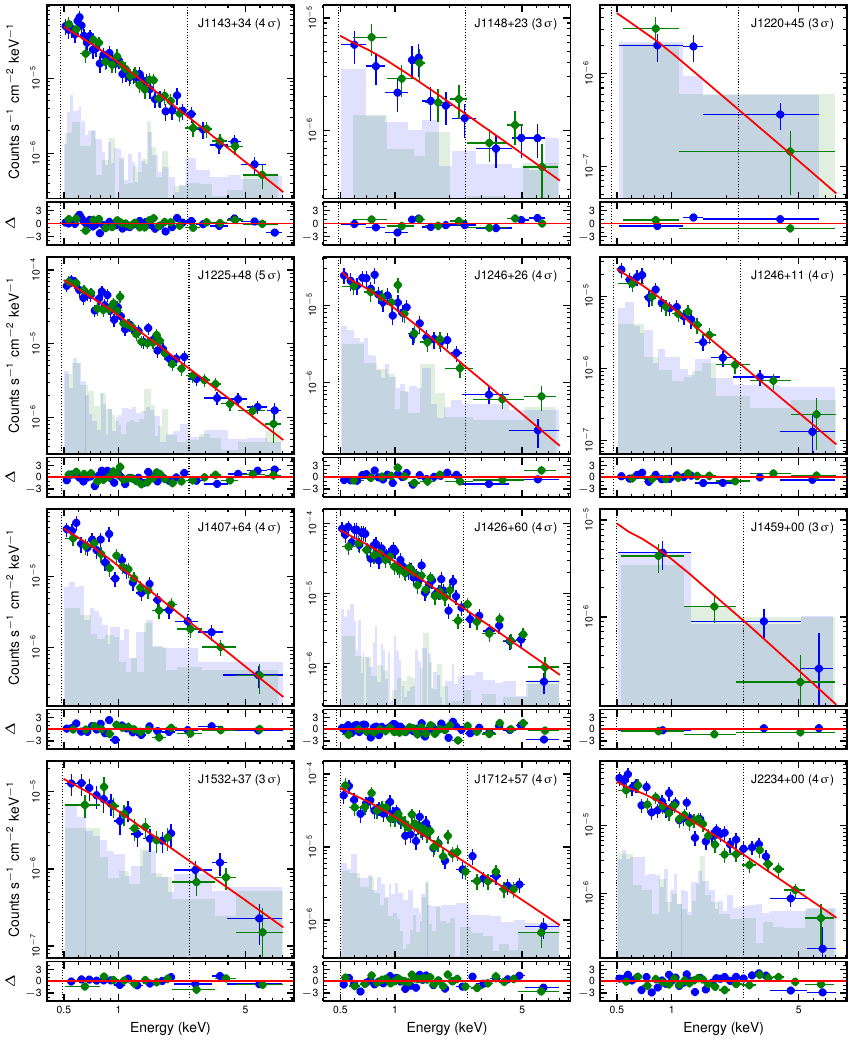}
\caption{continued.}
\end{figure*}

\section{Notes on individual objects}
\label{apb}

As per their nature as cosmic beacons, many of the 30 sources have been targeted with the most advanced facilities to address some of the hottest quasar-related astrophysical topics regarding the early Universe, among which are the  proximity effect and reionization, environment properties, and cosmic metallicity evolution. We report below some findings that might be relevant to the results discussed in this paper. \\
{\it J0304$-$00:} Recent observations with the integral field spectrograph MUSE at the Very Large Telescope have revealed a possible companion at a projected separation of only 20 kpc from the quasar \citep{Husemann+18}. This source is characterised by emission-line ratios consistent with photoionization from an AGN, yet likely obscured and less luminous than J0304$-$00 by three orders of magnitude. Hence, this satellite galaxy is not expected to contribute to the detected X-ray flux.  \\
{\it J0835+21:} This source was included in a sample of \civ BALs by \citet{Dunn+12}. Even so, the SDSS spectrum has a regular blue continuum, and the X-ray flux is such that $\daox = 0.03$. \\
{\it J0947+14:} This is one of the quasars shared with the WISSH sample, which has been recently classified as a BAL by \citet{Bruni+19} based on a double-dip absorption trough bluewards of the \siiv emission line. Due to the lack of obvious counterparts, this is interpreted as an ultra-fast \civ BAL with maximum outflow velocity of 0.15$c$. As for the previous source, however, the UV continuum is plainly blue and $\daox$ is positive. \\
{\it J1220+45:} A \civ, \siiv, \siv mini-BAL (i.e. with width $<$\,2000 \kms) system has been identifed by \citet{arav+18}. The ratio between the column densities of excited (S\,\textsc{iv}$^*$) and resonance (\siv) states places the absorbing gas at a distance of several hundreds of pc from the nucleus. Any association with the X-ray weakness appears rather challenging in the standard paradigm. \\
{\it J1425+54:} A claim for a \civ BAL with absorption index ${\rm AI}\simeq269$ \kms was made by \citet{Bruni+14}. The feature seems to be actually resolved in three narrow components, and its origin remains somewhat uncertain. \\
{\it J1426+60:} This source was analysed within a sample of 14 \civ mini-BALs with X-ray data by \citet{Wu+10}, who concluded that the quality of mini-BAL has no influence on the X-ray properties. Indeed, J1426+60 is the most X-ray luminous, radio-quiet quasar in our sample, with $\daox = 0.04$.  \\
{\it J1507+24:} A \civ BAL with minimum--maximum velocities of 18--$21 \times 10^3$ \kms and balnicity index ${\rm BI}\simeq150$ \kms was reported in this source by \citet{allen+11}.

\end{appendix}

%
%


\bibliographystyle{aa}
\bibliography{bibz3} 

\begin{thebibliography}{77}
\expandafter\ifx\csname natexlab\endcsname\relax\def\natexlab#1{#1}\fi

\bibitem[{{Allen} {et~al.}(2011){Allen}, {Hewett}, {Maddox}, {Richards}, \&
  {Belokurov}}]{allen+11}
{Allen}, J.~T., {Hewett}, P.~C., {Maddox}, N., {Richards}, G.~T., \&
  {Belokurov}, V. 2011, \mnras, 410, 860

\bibitem[{{Arav} {et~al.}(2018){Arav}, {Liu}, {Xu}, {Stidham}, {Benn}, \&
  {Chamberlain}}]{arav+18}
{Arav}, N., {Liu}, G., {Xu}, X., {et~al.} 2018, \apj, 857, 60

\bibitem[{{Avni} \& {Tananbaum}(1986)}]{Avni-Tananbaum_86}
{Avni}, Y. \& {Tananbaum}, H. 1986, \apj, 305, 83

\bibitem[{{Ba{\~n}ados} {et~al.}(2018){Ba{\~n}ados}, {Connor}, {Stern},
  {Mulchaey}, {Fan}, {Decarli}, {Farina}, {Mazzucchelli}, {Venemans}, {Walter},
  {Wang}, \& {Yang}}]{Banados+18}
{Ba{\~n}ados}, E., {Connor}, T., {Stern}, D., {et~al.} 2018, \apjl, 856, L25

\bibitem[{{Bahcall} \& {Hills}(1973)}]{Bahcall-Hills_73}
{Bahcall}, J.~N. \& {Hills}, R.~E. 1973, \apj, 179, 699

\bibitem[{{Baldwin}(1977)}]{Baldwin77}
{Baldwin}, J.~A. 1977, \apj, 214, 679

\bibitem[{{Baldwin} {et~al.}(1978){Baldwin}, {Burke}, {Gaskell}, \&
  {Wampler}}]{Baldwin+78}
{Baldwin}, J.~A., {Burke}, W.~L., {Gaskell}, C.~M., \& {Wampler}, E.~J. 1978,
  \nat, 273, 431

\bibitem[{{Bianchi} {et~al.}(2007){Bianchi}, {Guainazzi}, {Matt}, \& {Fonseca
  Bonilla}}]{Bianchi+07}
{Bianchi}, S., {Guainazzi}, M., {Matt}, G., \& {Fonseca Bonilla}, N. 2007,
  \aap, 467, L19

\bibitem[{{Bianchi} {et~al.}(2009){Bianchi}, {Guainazzi}, {Matt}, {Fonseca
  Bonilla}, \& {Ponti}}]{Bianchi+09}
{Bianchi}, S., {Guainazzi}, M., {Matt}, G., {Fonseca Bonilla}, N., \& {Ponti},
  G. 2009, \aap, 495, 421

\bibitem[{{Bischetti} {et~al.}(2017){Bischetti}, {Piconcelli}, {Vietri},
  {Bongiorno}, {Fiore}, {Sani}, {Marconi}, {Duras}, {Zappacosta}, {Brusa},
  {Comastri}, {Cresci}, {Feruglio}, {Giallongo}, {La Franca}, {Mainieri},
  {Mannucci}, {Martocchia}, {Ricci}, {Schneider}, {Testa}, \&
  {Vignali}}]{Bischetti+17}
{Bischetti}, M., {Piconcelli}, E., {Vietri}, G., {et~al.} 2017, \aap, 598, A122

\bibitem[{{Blomqvist} {et~al.}(2019){Blomqvist}, {du Mas des Bourboux},
  {Busca}, {de Sainte Agathe}, {Rich}, {Balland}, {Bautista}, {Dawson},
  {Font-Ribera}, {Guy}, {Le Goff}, {Palanque-Delabrouille}, {Percival},
  {P{\'e}rez-R{\`a}fols}, {Pieri}, {Schneider}, {Slosar}, \&
  {Y{\`e}che}}]{Blomqvist+19}
{Blomqvist}, M., {du Mas des Bourboux}, H., {Busca}, N.~G., {et~al.} 2019,
  \aap, 629, A86

\bibitem[{{Bruni} {et~al.}(2014){Bruni}, {Gonz{\'a}lez-Serrano}, {Pedani},
  {Benn}, {Mack}, {Holt}, {Montenegro-Montes}, \&
  {Jim{\'e}nez-Luj{\'a}n}}]{Bruni+14}
{Bruni}, G., {Gonz{\'a}lez-Serrano}, J.~I., {Pedani}, M., {et~al.} 2014, \aap,
  569, A87

\bibitem[{{Bruni} {et~al.}(2019){Bruni}, {Piconcelli}, {Misawa}, {Zappacosta},
  {Saturni}, {Vietri}, {Vignali}, {Bongiorno}, {Duras}, {Feruglio}, {Tombesi},
  \& {Fiore}}]{Bruni+19}
{Bruni}, G., {Piconcelli}, E., {Misawa}, T., {et~al.} 2019, \aap, 630, A111

\bibitem[{{Capellupo} {et~al.}(2016){Capellupo}, {Netzer}, {Lira},
  {Trakhtenbrot}, \& {Mej{\'\i}a-Restrepo}}]{Capellupo+16}
{Capellupo}, D.~M., {Netzer}, H., {Lira}, P., {Trakhtenbrot}, B., \&
  {Mej{\'\i}a-Restrepo}, J. 2016, \mnras, 460, 212

\bibitem[{{Cash}(1979)}]{Cash79}
{Cash}, W. 1979, \apj, 228, 939

\bibitem[{{Collinson} {et~al.}(2017){Collinson}, {Ward}, {Landt}, {Done},
  {Elvis}, \& {McDowell}}]{Collinson+17}
{Collinson}, J.~S., {Ward}, M.~J., {Landt}, H., {et~al.} 2017, \mnras, 465, 358

\bibitem[{{Czerny} \& {Elvis}(1987)}]{Czerny-Elvis_87}
{Czerny}, B. \& {Elvis}, M. 1987, \apj, 321, 305

\bibitem[{{Dunn} {et~al.}(2012){Dunn}, {Arav}, {Aoki}, {Wilkins}, {Laughlin},
  {Edmonds}, \& {Bautista}}]{Dunn+12}
{Dunn}, J.~P., {Arav}, N., {Aoki}, K., {et~al.} 2012, \apj, 750, 143

\bibitem[{{Elvis} \& {Karovska}(2002)}]{Elvis-Karovska_02}
{Elvis}, M. \& {Karovska}, M. 2002, \apjl, 581, L67

\bibitem[{{Fitzpatrick}(1999)}]{Fitzpatrick99}
{Fitzpatrick}, E.~L. 1999, \pasp, 111, 63

\bibitem[{{Gibson} {et~al.}(2008){Gibson}, {Brandt}, \&
  {Schneider}}]{Gibson+08}
{Gibson}, R.~R., {Brandt}, W.~N., \& {Schneider}, D.~P. 2008, \apj, 685, 773

\bibitem[{{Gibson} {et~al.}(2009){Gibson}, {Jiang}, {Brandt}, {Hall}, {Shen},
  {Wu}, {Anderson}, {Schneider}, {Vanden Berk}, {Gallagher}, {Fan}, \&
  {York}}]{Gibson+09}
{Gibson}, R.~R., {Jiang}, L., {Brandt}, W.~N., {et~al.} 2009, \apj, 692, 758

\bibitem[{{Haardt} \& {Maraschi}(1993)}]{Haardt-Maraschi_93}
{Haardt}, F. \& {Maraschi}, L. 1993, The Astrophysical Journal, 413, 507

\bibitem[{{Hewett} \& {Wild}(2010)}]{Hewett-Wild_10}
{Hewett}, P.~C. \& {Wild}, V. 2010, \mnras, 405, 2302

\bibitem[{{Hunter}(2007)}]{Hunter_07}
{Hunter}, J.~D. 2007, Computing in Science and Engineering, 9, 90

\bibitem[{{Husemann} {et~al.}(2018){Husemann}, {Worseck}, {Arrigoni Battaia},
  \& {Shanks}}]{Husemann+18}
{Husemann}, B., {Worseck}, G., {Arrigoni Battaia}, F., \& {Shanks}, T. 2018,
  \aap, 610, L7

\bibitem[{{Iwasawa} \& {Taniguchi}(1993)}]{Iwasawa-Taniguchi_93}
{Iwasawa}, K. \& {Taniguchi}, Y. 1993, \apjl, 413, L15

\bibitem[{{Just} {et~al.}(2007){Just}, {Brandt}, {Shemmer}, {Steffen},
  {Schneider}, {Chartas}, \& {Garmire}}]{Just+07}
{Just}, D.~W., {Brandt}, W.~N., {Shemmer}, O., {et~al.} 2007, \apj, 665, 1004

\bibitem[{{Kaastra}(2017)}]{Kaastra17}
{Kaastra}, J.~S. 2017, \aap, 605, A51

\bibitem[{{Kalberla} {et~al.}(2005){Kalberla}, {Burton}, {Hartmann}, {Arnal},
  {Bajaja}, {Morras}, \& {P{\"o}ppel}}]{Kalberla+05}
{Kalberla}, P.~M.~W., {Burton}, W.~B., {Hartmann}, D., {et~al.} 2005, \aap,
  440, 775

\bibitem[{{La Franca} {et~al.}(2014){La Franca}, {Bianchi}, {Ponti},
  {Branchini}, \& {Matt}}]{LaFranca+14}
{La Franca}, F., {Bianchi}, S., {Ponti}, G., {Branchini}, E., \& {Matt}, G.
  2014, \apjl, 787, L12

\bibitem[{{Laor} \& {Davis}(2014)}]{Laor-Davis_14}
{Laor}, A. \& {Davis}, S.~W. 2014, \mnras, 438, 3024

\bibitem[{{Laor} {et~al.}(1997){Laor}, {Fiore}, {Elvis}, {Wilkes}, \&
  {McDowell}}]{laor+97}
{Laor}, A., {Fiore}, F., {Elvis}, M., {Wilkes}, B.~J., \& {McDowell}, J.~C.
  1997, \apj, 477, 93

\bibitem[{{Leighly} {et~al.}(2007){Leighly}, {Halpern}, {Jenkins}, {Grupe},
  {Choi}, \& {Prescott}}]{Leighly+07}
{Leighly}, K.~M., {Halpern}, J.~P., {Jenkins}, E.~B., {et~al.} 2007, \apj, 663,
  103

\bibitem[{{Lightman} \& {Zdziarski}(1987)}]{Lightman-Zdziarski_87}
{Lightman}, A.~P. \& {Zdziarski}, A.~A. 1987, The Astrophysical Journal, 319,
  643

\bibitem[{{Liu} {et~al.}(2018){Liu}, {Luo}, {Brandt}, {Gallagher}, \&
  {Garmire}}]{Liu+18}
{Liu}, H., {Luo}, B., {Brandt}, W.~N., {Gallagher}, S.~C., \& {Garmire}, G.~P.
  2018, \apj, 859, 113

\bibitem[{{Luo} {et~al.}(2014){Luo}, {Brandt}, {Alexander}, {Stern}, {Teng},
  {Ar{\'e}valo}, {Bauer}, {Boggs}, {Christensen}, {Comastri}, {Craig},
  {Farrah}, {Gandhi}, {Hailey}, {Harrison}, {Koss}, {Ogle}, {Puccetti}, {Saez},
  {Scott}, {Walton}, \& {Zhang}}]{Luo+14}
{Luo}, B., {Brandt}, W.~N., {Alexander}, D.~M., {et~al.} 2014, \apj, 794, 70

\bibitem[{{Luo} {et~al.}(2015){Luo}, {Brandt}, {Hall}, {Wu}, {Anderson},
  {Garmire}, {Gibson}, {Plotkin}, {Richards}, {Schneider}, {Shemmer}, \&
  {Shen}}]{Luo+15}
{Luo}, B., {Brandt}, W.~N., {Hall}, P.~B., {et~al.} 2015, \apj, 805, 122

\bibitem[{{Lusso} {et~al.}(2010){Lusso}, {Comastri}, {Vignali}, {Zamorani},
  {Brusa}, {Gilli}, {Iwasawa}, {Salvato}, {Civano}, {Elvis}, {Merloni},
  {Bongiorno}, {Trump}, {Koekemoer}, {Schinnerer}, {Le Floc'h}, {Cappelluti},
  {Jahnke}, {Sargent}, {Silverman}, {Mainieri}, {Fiore}, {Bolzonella}, {Le
  F{\`e}vre}, {Garilli}, {Iovino}, {Kneib}, {Lamareille}, {Lilly}, {Mignoli},
  {Scodeggio}, \& {Vergani}}]{Lusso+10}
{Lusso}, E., {Comastri}, A., {Vignali}, C., {et~al.} 2010, \aap, 512, A34

\bibitem[{{Lusso} {et~al.}(2013){Lusso}, {Hennawi}, {Comastri}, {Zamorani},
  {Richards}, {Vignali}, {Treister}, {Schawinski}, {Salvato}, \&
  {Gilli}}]{Lusso+13}
{Lusso}, E., {Hennawi}, J.~F., {Comastri}, A., {et~al.} 2013, \apj, 777, 86

\bibitem[{{Lusso} \& {Risaliti}(2016)}]{LR16}
{Lusso}, E. \& {Risaliti}, G. 2016, \apj, 819, 154

\bibitem[{{Lusso} \& {Risaliti}(2017)}]{LR17}
{Lusso}, E. \& {Risaliti}, G. 2017, \aap, 602, A79

\bibitem[{{Lusso} {et~al.}(2015){Lusso}, {Worseck}, {Hennawi}, {Prochaska},
  {Vignali}, {Stern}, \& {O'Meara}}]{Lusso+15}
{Lusso}, E., {Worseck}, G., {Hennawi}, J.~F., {et~al.} 2015, \mnras, 449, 4204

\bibitem[{{Martocchia} {et~al.}(2017){Martocchia}, {Piconcelli}, {Zappacosta},
  {Duras}, {Vietri}, {Vignali}, {Bianchi}, {Bischetti}, {Bongiorno}, {Brusa},
  {Lanzuisi}, {Marconi}, {Mathur}, {Miniutti}, {Nicastro}, {Bruni}, \&
  {Fiore}}]{Martocchia+17}
{Martocchia}, S., {Piconcelli}, E., {Zappacosta}, L., {et~al.} 2017, \aap, 608,
  A51

\bibitem[{{Marziani} \& {Sulentic}(2014)}]{Marziani-Sulentic_14}
{Marziani}, P. \& {Sulentic}, J.~W. 2014, \mnras, 442, 1211

\bibitem[{{Mingo} {et~al.}(2016){Mingo}, {Watson}, {Rosen}, {Hardcastle},
  {Ruiz}, {Blain}, {Carrera}, {Mateos}, {Pineau}, \& {Stewart}}]{Mingo+16}
{Mingo}, B., {Watson}, M.~G., {Rosen}, S.~R., {et~al.} 2016, \mnras, 462, 2631

\bibitem[{{Mosquera} {et~al.}(2013){Mosquera}, {Kochanek}, {Chen}, {Dai},
  {Blackburne}, \& {Chartas}}]{Mosquera+13}
{Mosquera}, A.~M., {Kochanek}, C.~S., {Chen}, B., {et~al.} 2013, \apj, 769, 53

\bibitem[{{Murray} {et~al.}(1995){Murray}, {Chiang}, {Grossman}, \&
  {Voit}}]{Murray+95}
{Murray}, N., {Chiang}, J., {Grossman}, S.~A., \& {Voit}, G.~M. 1995, \apj,
  451, 498

\bibitem[{{Nanni} {et~al.}(2018){Nanni}, {Gilli}, {Vignali}, {Mignoli},
  {Comastri}, {Vanzella}, {Zamorani}, {Calura}, {Lanzuisi}, {Brusa}, {Tozzi},
  {Iwasawa}, {Cappi}, {Vito}, {Balmaverde}, {Costa}, {Risaliti}, {Paolillo},
  {Prandoni}, {Liuzzo}, {Rosati}, {Chiaberge}, {Caminha}, {Sani}, {Cappelluti},
  \& {Norman}}]{Nanni+18}
{Nanni}, R., {Gilli}, R., {Vignali}, C., {et~al.} 2018, \aap, 614, A121

\bibitem[{{Nardini} {et~al.}(2015){Nardini}, {Reeves}, {Gofford}, {Harrison},
  {Risaliti}, {Braito}, {Costa}, {Matzeu}, {Walton}, {Behar}, {Boggs},
  {Christensen}, {Craig}, {Hailey}, {Matt}, {Miller}, {O'Brien}, {Stern},
  {Turner}, \& {Ward}}]{Nardini+15}
{Nardini}, E., {Reeves}, J.~N., {Gofford}, J., {et~al.} 2015, Science, 347, 860

\bibitem[{{Ni} {et~al.}(2018){Ni}, {Brandt}, {Luo}, {Hall}, {Shen}, {Anderson},
  {Plotkin}, {Richards}, {Schneider}, {Shemmer}, \& {Wu}}]{Ni+18}
{Ni}, Q., {Brandt}, W.~N., {Luo}, B., {et~al.} 2018, \mnras, 480, 5184

\bibitem[{{P{\^a}ris} {et~al.}(2018){P{\^a}ris}, {Petitjean}, {Aubourg},
  {Myers}, {Streblyanska}, {Lyke}, {Anderson}, {Armengaud}, {Bautista},
  {Blanton}, {Blomqvist}, {Brinkmann}, {Brownstein}, {Brand t}, {Burtin},
  {Dawson}, {de la Torre}, {Georgakakis}, {Gil-Mar{\'\i}n}, {Green}, {Hall},
  {Kneib}, {LaMassa}, {Le Goff}, {MacLeod}, {Mariappan}, {McGreer}, {Merloni},
  {Noterdaeme}, {Palanque-Delabrouille}, {Percival}, {Ross}, {Rossi},
  {Schneider}, {Seo}, {Tojeiro}, {Weaver}, {Weijmans}, {Y{\`e}che}, {Zarrouk},
  \& {Zhao}}]{Paris+18}
{P{\^a}ris}, I., {Petitjean}, P., {Aubourg}, {\'E}., {et~al.} 2018, \aap, 613,
  A51

\bibitem[{{Park} {et~al.}(2006){Park}, {Kashyap}, {Siemiginowska}, {van Dyk},
  {Zezas}, {Heinke}, \& {Wargelin}}]{Park+06}
{Park}, T., {Kashyap}, V.~L., {Siemiginowska}, A., {et~al.} 2006, \apj, 652,
  610

\bibitem[{{Piconcelli} {et~al.}(2005){Piconcelli}, {Jimenez-Bail{\'o}n},
  {Guainazzi}, {Schartel}, {Rodr{\'\i}guez-Pascual}, \&
  {Santos-Lle{\'o}}}]{Piconcelli+05}
{Piconcelli}, E., {Jimenez-Bail{\'o}n}, E., {Guainazzi}, M., {et~al.} 2005,
  \aap, 432, 15

\bibitem[{{Proga}(2005)}]{Proga_05}
{Proga}, D. 2005, \apjl, 630, L9

\bibitem[{{Richards} {et~al.}(2006){Richards}, {Lacy}, {Storrie-Lombardi},
  {Hall}, {Gallagher}, {Hines}, {Fan}, {Papovich}, {Vanden Berk}, {Trammell},
  {Schneider}, {Vestergaard}, {York}, {Jester}, {Anderson}, {Budav{\'a}ri}, \&
  {Szalay}}]{Richards+06}
{Richards}, G.~T., {Lacy}, M., {Storrie-Lombardi}, L.~J., {et~al.} 2006, \apjs,
  166, 470

\bibitem[{{Risaliti} \& {Lusso}(2015)}]{RL15}
{Risaliti}, G. \& {Lusso}, E. 2015, \apj, 815, 33

\bibitem[{{Risaliti} \& {Lusso}(2019)}]{RL19}
{Risaliti}, G. \& {Lusso}, E. 2019, Nature Astronomy, 195

\bibitem[{{Rosen} {et~al.}(2016){Rosen}, {Webb}, {Watson}, {Ballet}, {Barret},
  {Braito}, {Carrera}, {Ceballos}, {Coriat}, {Della Ceca}, {Denkinson},
  {Esquej}, {Farrell}, {Freyberg}, {Gris{\'e}}, {Guillout}, {Heil},
  {Koliopanos}, {Law-Green}, {Lamer}, {Lin}, {Martino}, {Michel}, {Motch},
  {Nebot Gomez-Moran}, {Page}, {Page}, {Page}, {Pakull}, {Pye}, {Read},
  {Rodriguez}, {Sakano}, {Saxton}, {Schwope}, {Scott}, {Sturm}, {Traulsen},
  {Yershov}, \& {Zolotukhin}}]{Rosen+16}
{Rosen}, S.~R., {Webb}, N.~A., {Watson}, M.~G., {et~al.} 2016, \aap, 590, A1

\bibitem[{{Schlegel} {et~al.}(1998){Schlegel}, {Finkbeiner}, \&
  {Davis}}]{Schlegel+98}
{Schlegel}, D.~J., {Finkbeiner}, D.~P., \& {Davis}, M. 1998, \apj, 500, 525

\bibitem[{{Scolnic} {et~al.}(2018){Scolnic}, {Jones}, {Rest}, {Pan},
  {Chornock}, {Foley}, {Huber}, {Kessler}, {Narayan}, {Riess}, {Rodney},
  {Berger}, {Brout}, {Challis}, {Drout}, {Finkbeiner}, {Lunnan}, {Kirshner},
  {Sand ers}, {Schlafly}, {Smartt}, {Stubbs}, {Tonry}, {Wood-Vasey}, {Foley},
  {Hand}, {Johnson}, {Burgett}, {Chambers}, {Draper}, {Hodapp}, {Kaiser},
  {Kudritzki}, {Magnier}, {Metcalfe}, {Bresolin}, {Gall}, {Kotak}, {McCrum}, \&
  {Smith}}]{Scolnic+18}
{Scolnic}, D.~M., {Jones}, D.~O., {Rest}, A., {et~al.} 2018, \apj, 859, 101

\bibitem[{{Setti} \& {Woltjer}(1973)}]{Setti-Woltjer_73}
{Setti}, G. \& {Woltjer}, L. 1973, \apjl, 181, L61

\bibitem[{{Shakura} \& {Sunyaev}(1973)}]{SS73}
{Shakura}, N.~I. \& {Sunyaev}, R.~A. 1973, \aap, 24, 337

\bibitem[{{Shemmer} {et~al.}(2017){Shemmer}, {Brandt}, {Paolillo}, {Kaspi},
  {Vignali}, {Lira}, \& {Schneider}}]{Shemmer+17}
{Shemmer}, O., {Brandt}, W.~N., {Paolillo}, M., {et~al.} 2017, \apj, 848, 46

\bibitem[{{Shen} {et~al.}(2011){Shen}, {Richards}, {Strauss}, {Hall},
  {Schneider}, {Snedden}, {Bizyaev}, {Brewington}, {Malanushenko},
  {Malanushenko}, {Oravetz}, {Pan}, \& {Simmons}}]{Shen+11}
{Shen}, Y., {Richards}, G.~T., {Strauss}, M.~A., {et~al.} 2011, \apjs, 194, 45

\bibitem[{{Slone} \& {Netzer}(2012)}]{Slone-Netzer_12}
{Slone}, O. \& {Netzer}, H. 2012, \mnras, 426, 656

\bibitem[{{Steffen} {et~al.}(2006){Steffen}, {Strateva}, {Brandt}, {Alexand
  er}, {Koekemoer}, {Lehmer}, {Schneider}, \& {Vignali}}]{Steffen+06}
{Steffen}, A.~T., {Strateva}, I., {Brandt}, W.~N., {et~al.} 2006, \aj, 131,
  2826

\bibitem[{{Strateva} {et~al.}(2005){Strateva}, {Brandt}, {Schneider}, {Vanden
  Berk}, \& {Vignali}}]{Strateva+05}
{Strateva}, I.~V., {Brandt}, W.~N., {Schneider}, D.~P., {Vanden Berk}, D.~G.,
  \& {Vignali}, C. 2005, \aj, 130, 387

\bibitem[{{Taylor}(2005)}]{Taylor_05}
{Taylor}, M.~B. 2005, in Astronomical Society of the Pacific Conference Series,
  Vol. 347, Astronomical Data Analysis Software and Systems XIV, ed.
  P.~{Shopbell}, M.~{Britton}, \& R.~{Ebert}, 29

\bibitem[{{Teng} {et~al.}(2014){Teng}, {Brandt}, {Harrison}, {Luo},
  {Alexander}, {Bauer}, {Boggs}, {Christensen}, {Comastri}, {Craig}, {Fabian},
  {Farrah}, {Fiore}, {Gandhi}, {Grefenstette}, {Hailey}, {Hickox}, {Madsen},
  {Ptak}, {Rigby}, {Risaliti}, {Saez}, {Stern}, {Veilleux}, {Walton}, {Wik}, \&
  {Zhang}}]{Teng+14}
{Teng}, S.~H., {Brandt}, W.~N., {Harrison}, F.~A., {et~al.} 2014, \apj, 785, 19

\bibitem[{{Vanden Berk} {et~al.}(2001){Vanden Berk}, {Richards}, {Bauer},
  {Strauss}, {Schneider}, {Heckman}, {York}, {Hall}, {Fan}, {Knapp},
  {Anderson}, {Annis}, {Bahcall}, {Bernardi}, {Briggs}, {Brinkmann}, {Brunner},
  {Burles}, {Carey}, {Castander}, {Connolly}, {Crocker}, {Csabai}, {Doi},
  {Finkbeiner}, {Friedman}, {Frieman}, {Fukugita}, {Gunn}, {Hennessy},
  {Ivezi{\'c}}, {Kent}, {Kunszt}, {Lamb}, {Leger}, {Long}, {Loveday}, {Lupton},
  {Meiksin}, {Merelli}, {Munn}, {Newberg}, {Newcomb}, {Nichol}, {Owen}, {Pier},
  {Pope}, {Rockosi}, {Schlegel}, {Siegmund}, {Smee}, {Snir}, {Stoughton},
  {Stubbs}, {SubbaRao}, {Szalay}, {Szokoly}, {Tremonti}, {Uomoto}, {Waddell},
  {Yanny}, \& {Zheng}}]{VandenBerk+01}
{Vanden Berk}, D.~E., {Richards}, G.~T., {Bauer}, A., {et~al.} 2001, \aj, 122,
  549

\bibitem[{{Vignali} {et~al.}(2003){Vignali}, {Brandt}, \&
  {Schneider}}]{Vignali+03}
{Vignali}, C., {Brandt}, W.~N., \& {Schneider}, D.~P. 2003, \aj, 125, 433

\bibitem[{{Vito} {et~al.}(2019){Vito}, {Brandt}, {Bauer}, {Calura}, {Gilli},
  {Luo}, {Shemmer}, {Vignali}, {Zamorani}, {Brusa}, {Civano}, {Comastri}, \&
  {Nanni}}]{Vito+19}
{Vito}, F., {Brandt}, W.~N., {Bauer}, F.~E., {et~al.} 2019, \aap, 630, A118

\bibitem[{{Wang} {et~al.}(2013){Wang}, {Du}, {Valls-Gabaud}, {Hu}, \&
  {Netzer}}]{Wang+13}
{Wang}, J.-M., {Du}, P., {Valls-Gabaud}, D., {Hu}, C., \& {Netzer}, H. 2013,
  \prl, 110, 081301

\bibitem[{{Watson} {et~al.}(2011){Watson}, {Denney}, {Vestergaard}, \&
  {Davis}}]{Watson+11}
{Watson}, D., {Denney}, K.~D., {Vestergaard}, M., \& {Davis}, T.~M. 2011,
  \apjl, 740, L49

\bibitem[{{Weisskopf} {et~al.}(2007){Weisskopf}, {Wu}, {Trimble}, {O'Dell},
  {Elsner}, {Zavlin}, \& {Kouveliotou}}]{Weisskopf+07}
{Weisskopf}, M.~C., {Wu}, K., {Trimble}, V., {et~al.} 2007, \apj, 657, 1026

\bibitem[{{Wu} {et~al.}(2010){Wu}, {Brandt}, {Comins}, {Gibson}, {Shemmer},
  {Garmire}, \& {Schneider}}]{Wu+10}
{Wu}, J., {Brandt}, W.~N., {Comins}, M.~L., {et~al.} 2010, \apj, 724, 762

\end{thebibliography}

\end{document}